\documentclass[10pt]{article}

\usepackage{graphicx}
\usepackage{epsfig}
\usepackage{amsmath}
\usepackage{amssymb}

\newcommand{\TIdiff}{\Delta\,                      
\hat C[\sigma^{(a)}_\tot , \sigma^{(b)}_\tot](t) / T^2 }
\newcommand{\TdAuto}{\Delta\,                      
\hat C[\sigma_\tot](t) / T^2 }

\newcommand{\intern}{{\mbox{\tiny int}}}
\newcommand{\pint}{-\!\!\!\!\!\!\!\int}
\newcommand{\rmi}{{\rm i}}
\newcommand{\rmd}{{\rm d}}
\newcommand{\rme}{{\rm e}}
\newcommand{\la}{\langle}
\newcommand{\ra}{\rangle}
\newcommand{\lla}{\left\langle}
\newcommand{\rra}{\right\rangle}
\newcommand{\tot}{{\mbox{\tiny tot}}}
\newcommand{\eps}{\varepsilon}

\begin{document}

\title{Signatures of the correlation hole in total and partial cross
sections}

\author{T. Gorin\footnote{
Present address: Theoretical Quantumdynamics,
Albert-Ludwigs-Universit\" at,
Hermann-Herder Str. 3, 79104 Freiburg.
E-mail: Thomas.Gorin@physik.uni-freiburg.de}
\ and T. H. Seligman\\
{\small Centro de Ciencias F\'\i sicas, University of Mexico (UNAM), C. P. 62210
Cuernavaca, M\'exico}}

\date{}

\maketitle

\begin{abstract}
In a complex scattering system with few open channels, say a quantum
dot with leads, the correlation properties of the poles of the
scattering matrix are most directly related to the internal dynamics of the
system. We may ask how to extract these properties from an analysis of cross 
sections. In general this is very difficult, if we leave the domain of isolated 
resonances.  We propose to consider the cross correlation function of two 
different elastic or total cross sections. For these we can show numerically 
and to some extent also analytically a significant dependence on the 
correlations between the scattering poles. The difference between uncorrelated 
and strongly correlated poles is clearly visible, even for strongly overlapping 
resonances. \\

\noindent
PACS number(s): 05.45.Mt, 
                03.65.Nk  
\end{abstract}

\section{Introduction \label{sI}}

Starting from Bohr's compound nucleus suggestion, the idea of considering the 
dynamics in the interaction region almost separately from the scattering 
process has been proven very successful in different fields. Wigner's 
R-matrix method \cite{WigEis47} gives the formal background to the separation 
of internal dynamics and ``free'' motion in the channel space. Based on
this idea we study whether chaoticity or integrability of the internal 
dynamics can be detected in the scattering data. For this purpose we apply 
Fourier transform techniques, which have proved successful in spectral analysis 
\cite{Lev86,Guh90,Lom93}, to total and partial cross sections.

To build the scattering ensembles, we shall assume that there are no 
correlations between channel space and the internal structure. This assumption 
is usually well fulfilled for systems with topological chaos, while it is 
often not fulfilled for integrable systems \cite{JungSeligman97}. Nevertheless,
we use this assumption in order to compare the chaotic and regular case in a 
direct and minimally biased way. Any differences we then find are basis 
independent and minimal in the sense that correlations would usually increase 
the dissimilarity to the chaotic case. Therefore, we use orthogonally invariant 
random matrix models to describe the internal structure. For the chaotic case 
the choice is obviously the Gaussian orthogonal ensemble (GOE) if time reversal 
symmetry is conserved. Following Berry and Tabor \cite{BerTab77} we associate 
integrability with a random Poissonian spectrum, thus excluding harmonic 
oscillators explicitly. For this case the Poisson orthogonal ensemble (POE) was 
proposed some years ago \cite{Dit91b}. The case of time reversal symmetry 
breaking can be treated analogously using the unitary ensembles, but this will 
not be discussed in the present paper.

We consider three different absorption regimes. For {\em weak coupling} the 
resonances are isolated, and conventional spectral analysis is satisfactory; if 
we wish we may add an analysis of the widths. Then follows what is usually 
called {\em strong absorption}, where we find overlapping resonances, but where 
the individual transmissions from all channels are considerably smaller than 
one. Next we have the case where the transmissions are close to one. To 
differentiate the latter two regimes, we shall speak of {\em strong 
transmission} in this case. It corresponds to the semiclassical limit, where 
tunneling effects become negligible.

For level spectra as well as for intensity spectra, Fourier transform methods
and the so called ``correlation hole'' have been very successful tools to 
identify the effects of integrability and chaos 
\cite{Lev86,Guh90,Lom93,AlGuHa97}. Yet it is not clear to what extent such an
analysis remains useful, when the resonances are no longer isolated. An exact
theory is only available for two-point functions of S-matrix elements in the
GOE case. This so called Verbaarschot-Weidenm\" uller-Zirnbauer (VWZ) integral 
\cite{VWZ85} allows to calculate 
correlation functions of total cross sections by means of the optical theorem. 
In contrast to that, we have no such theory in the POE case or for correlation 
functions of partial cross sections. In these cases, one had to fall back on 
the so called ``Breit-Wigner'' approximation \cite{BreWig36}, which becomes 
valid in the weak coupling limit. To extend the validity of this 
approximation we take advantage of the well known fact that the Satchler 
transmission matrix \cite{Sat63} or, in absence of direct reactions as in our
case, the transmission coefficients are the only way in which the coupling 
strength influences the physically relevant quantities. 
By using the transmission coefficients rather than the coupling constants as 
input, which amounts to a rescaling, we are able to extend the validity 
of this approximation to the regime of {\em strong absorption}, 
where the resonances are overlapping, but the absorption in each channel is 
still moderate.

We perform numerical simulations with two purposes: to check the range of
validity of the rescaled Breit-Wigner formalism, and to find situations, 
which show significant differences between GOE and POE. We will concentrate on 
Fourier transforms of auto and cross correlation functions, and we shall see 
that in particular for the latter the differences are in some cases very 
promising. In particular we find that cross correlations between cross sections
from different channels show strong signatures.\\

In Sec.~\ref{M} we present the model we use. In Sec.~\ref{C} we 
introduce the correlation functions of S-matrix elements and cross sections, 
our basic tool for the statistical analysis of the scattering systems.
Next we discuss the ``rescaled Breit-Wigner'' approximation in Sec.~\ref{R}. 
It allows to get results in closed form for the correlation functions between 
total or partial cross sections. In Sec.~\ref{V} we derive a twofold
integral expression for  the Fourier transform of the VWZ integral 
\cite{VWZ85}. This simplifies the numerical treatment considerably, and it is 
used in Sec.~\ref{T}, to test the validity of the rescaled Breit-Wigner 
approximation. After these theoretical considerations, we turn to the numerical
study of the two scattering ensembles, the POE and  the GOE. This is done in 
Sec.~\ref{N}, which is divided into two subsections. The first deals with 
correlation functions between total cross sections, and the second with 
correlation functions between partial cross sections. Section~\ref{S} contains 
a short summary.

\section{The scattering ensembles \label{M}}

We wish to construct scattering ensembles for the two contrary cases, 
where the dynamics in the interaction region is predominantly integrable, or 
completely chaotic. We do this under the assumption that the scattering system 
may be separated into an internal part restricted to a finite interaction 
region and an external part described by some superintegrable Hamiltonian.
Though some complexity may show up in the coupling of the two parts also, 
{\it i.e.} in the mismatch of channel functions and internal functions, the
dominant part of the complexity should be contained in the internal part.
Then the subsystem that describes the complex internal
dynamics has a discrete spectrum, such that its statistical properties can be
modeled with an appropriate random matrix ensemble. In order to construct the
scattering ensembles, we fix the external part and use standard techniques,
originally introduced to describe nuclear compound reactions 
\cite{Feshbach58,Feshbach62,MahWei69}, to assign to each element from the 
random matrix ensemble a scattering matrix. Thus we obtain a set of scattering 
matrices provided with the measure inherited from the original random matrix 
ensemble. For simplicity we will denote the scattering ensembles obtained from
the GOE and the POE by the same names whenever there is no danger for 
confusion.

To be more precise, we consider complex scattering systems with many, 
possibly overlapping, resonances, where the S-matrix can be cast into the 
following form:
\begin{equation}
S(E) = 1 - \rmi \, V^\dagger \frac{1}{E-H} V \; , \quad
H = H_\intern - \frac{\rmi}{2} V V^\dagger  \; .
\label{M_Smat}\end{equation}
Here $H_\intern$ is a real, symmetric $N \times N$ matrix which describes the 
internal dynamics, and $V$ is a real $N \times M$ matrix, describing the 
coupling to the $M$ channels. The matrix $H$ is the so called effective 
Hamiltonian \cite{Kle85}. In order to arrive at Eq.~(\ref{M_Smat}) it is
assumed that the coupling matrix elements between channel states and internal 
eigenstates are energy independent \cite{MahWei69}. Furthermore one should 
either neglect the direct reactions or perform an Engelbrecht-Weidenm\"uller 
transformation \cite{EngWei73} if it is necessary to take them into 
account. The effective Hamiltonian $H$ can be diagonalized, such that its 
eigenvalues $\tilde E_j =  E_j -\rmi\Gamma_j/2$ give the positions and widths 
of the resonances if they are isolated. In the eigenbasis of $H$, the 
S-matrix elements can be written as
\begin{eqnarray}
S_{ab}(E) &=& \delta_{ab} - \rmi \sum_{j=1}^N
\frac{\tilde{V}_{ja} \tilde{V}_{jb}}{E - \tilde{E}_j} \; ,
\label{M_Smat2}\\
\tilde V &=& A^T V \; , \quad A^T H A = {\rm diag}\left(\tilde E_j\right) \; .
\nonumber\end{eqnarray}
This equation shows that the complex poles of the S-matrix are precisely the 
$N$ eigenvalues of $H$. If the coupling matrix elements are small enough, their 
real parts are well approximated by the discrete levels of $H_\intern$, whereas 
their imaginary parts are given by the diagonal elements of $VV^\dagger$. This 
amounts to the Breit-Wigner approximation, which results from applying first 
order perturbation theory to the effective Hamiltonian $H$. \\

In this paper we consider two scattering ensembles: the GOE and the POE.
Both are invariant with respect to orthogonal transformations. Hence in the
eigenbasis of $H_\intern$ the $M$ channel vectors are random orthogonal 
vectors. In practice, we use independent random vectors with Gaussian 
distributed components for $V$, which are orthogonal only up to order 
${\cal O}(N^{-1})$. However, as we used relatively large matrices $N=300$, the 
violation of the orthogonal invariance had certainly no noticeable effect on 
our numerical results. Hence, for both ensembles the nonzero eigenvalues of 
$VV^\dagger$ (given by the norm squared of the column vectors), and 
the level density of $H_\intern$ are the only independent parameters.

In the GOE case, the elements of the diagonal matrix $H_\intern$ are 
distributed according to the joint probability distribution of the GOE spectrum 
\cite{Mehta91}, so that for large $N$, the level density approaches the 
semicircle distribution. In the POE case, the elements are independently 
distributed, and in principle the level density can be of any form. Our main 
objective is the distinction between both ensembles, based on the observation 
of correlations. Hence we find it convenient to use the semicircle 
distribution in the POE case also. \\

The main theoretical tool for the calculation of correlation functions will be
the rescaled Breit-Wigner approximation, introduced in Sec.~\ref{R}. In fact,
it can be applied in much more general situations. This is of particular 
importance for partially integrable scattering systems, where the assumption of 
orthogonal invariance does often not hold. Then the distribution of the 
matrix elements of $V$ are typically very different from simple uncorrelated 
Gaussian distributions. \\

In cases as they are studied here, the openness of the scattering system is
commonly described, borrowing the terminology from the so called
``optical model'' ({\it cf.} \cite{MahWei69} and references therein), which was 
originally developed to describe nuclear compound reactions with two well 
separated time scales. Consider the partial cross section in appropriate units,
which is  given by: $\sigma_{ab} = |\delta_{ab} - S_{ab}|^2$. Then one defines 
the optical partial cross section as 
$\sigma_{ab}^{\mbox{\tiny opt}} = |\delta_{ab} - \la S_{ab}\ra|^2$, where
the different time scales are used to obtain a well defined average
S-matrix $\la S_{ab}\ra$ by averaging over an appropriately chosen energy
window. Here $\la S_{ab}\ra$ is simply defined by the ensemble average,
avoiding in this way any arbitrariness. The openness of the scattering system 
is then characterized by so called ``transmission coefficients,'' defined for 
each entrance channel $a$:
\begin{equation}
T_a = \lla\sigma_\tot^{(a)}\rra - \sigma_{\mbox{\tiny opt}}^{(a)} = 
1 - \sum_{c=1}^M |\la S_{ac}\ra|^2 \; ,
\label{M_trami}\end{equation}
where $\sigma_\tot^{(a)}= \sum_{c=1}^M \sigma_{ac}$ is the total cross section,
and $\sigma_{\mbox{\tiny opt}}^{(a)} = \sum_{c=1}^M 
\sigma_{ac}^{\mbox{\tiny opt}}$ is the total optical cross section, with 
respect to the entrance channel $a$. The unitarity of the S-matrix leads to:
\begin{equation}
\sigma_\tot^{(a)} = 2\left( 1 - {\rm Re}\, S_{aa}\right) \; ,
\label{M_optthm}\end{equation}
which is sometimes called the ``optical theorem.''

The scattering ensembles defined above, {\it i.e.} the GOE and the POE, are 
completely characterized by the average level distance $d= (N\rho)^{-1}$ in
the center of the spectrum of $H_\intern$, and the variance of the coupling 
matrix elements $\la V_{ia}^2\ra$, which are independent of $i$ due to
orthogonal invariance. From those we define the following dimensionless
coupling parameters:
\begin{equation}
\kappa_a = \frac{\pi}{2d}\la V_{ia}^2\ra \; .
\label{M_kap}\end{equation}
Here and in what follows, the brackets $\la \ldots\ra$ stand for the ensemble 
average. As discussed in the Appendix, the average S-matrix is diagonal.
Its diagonal elements, the transmission coefficients, and the coupling 
parameters are related to each other:
\begin{equation}
\la S_{aa}\ra = \frac{1-\kappa_a}{1+\kappa_a} \; \Leftrightarrow \;
T_a = \frac{4\kappa_a}{(1+\kappa_a)^2} \; ,
\label{M_Tau}\end{equation}
which is, however, true in the center of the spectrum only (a more detailed 
discussion is given in the Appendix).

As mentioned in the introduction, we distinguish three different regimes. Now,
these can be defined more quantitatively in terms of the transmission 
coefficients: The first is the {\em weak absorption} regime, where the resonances 
are still well separated, so that $\sum_{a=1}^M T_a \ll 1$. Next comes the regime 
of {\em strong absorption}, where the resonances overlap, but the transmission 
from each channel is still small: $\sum_{a=1}^M T_a > 1 ,\, \forall a: T_a \ll 1$. 
Note, that this implies in particular, that the number of channels $M$ is 
large. Finally we have the regime of {\em strong transmission}: 
$\forall a: T_a \lesssim 1$ where the transmissions in all channels are close 
to one.

\section{Correlation functions of total and partial cross sections \label{C}}

Correlation functions are our principal tool for the statistical analysis of
total and partial cross sections. We distinguish between autocorrelation 
functions, where one cross section is correlated with itself, and cross 
correlation functions, where two different cross sections are correlated with
each other. First we define the correlation functions in general, in order to
introduce our notation. Then we use the optical theorem, to relate the 
correlation function of two total cross sections to one of corresponding 
S-matrix elements. For the latter, the VWZ integral \cite{VWZ85} provides the
exact result in the GOE case. In the POE case an exact result exists only in 
the one channel case \cite{Gor99}, and we use the rescaled Breit-Wigner 
approximation there. For partial cross sections, no exact theory exists at all. 
In this case, we first use the so called ``diagonal approximation'' to express 
the partial cross sections in such a form, that the rescaled Breit-Wigner 
approximation can be applied.

Dealing with matrix ensembles, it is convenient to define the correlation 
functions as ensemble averages, rather then energy averages. Therefore, we 
eventually have to face the ergodicity question \cite{Pan79}, which is 
unclear in the POE case. Note, however, that in quantum dot experiments ensemble 
averages may actually be the relevant ones \cite{Chan95}. We calculate the 
correlation functions always in the center of the spectrum (see Appendix), where 
we set $E=0$. Given then two complex functions $f$ and $g$ of the energy, we 
define the correlation function as follows:
\begin{equation}
C[f,g](\omega) = \lla f\!\!\left(\textstyle \frac{-\omega d}{2}\right)\,
g\!\!\left(\textstyle\frac{\omega d}{2}\right)\rra - 
\lla f\!\!\left(\textstyle \frac{-\omega d}{2}\right)\rra \, 
\lla g\!\!\left(\textstyle\frac{\omega d}{2}\right) \rra \; .
\label{C_Cfg}\end{equation}
Here $d$ is the mean level distance in the center of the spectrum of 
$H_\intern$ which is assumed to be constant on the scale where we expect 
correlations. Note that, there is no unfolding involved. The mean level 
distance $d$ simply serves as a convenient energy scale. For the discrete 
spectrum of some random Hamiltonian: 
$f(E)= \sum_i \delta(E - E_i)$, the autocorrelation function becomes: 
$C[f,f^*](\omega) = 1+\delta(\omega) - Y_2(\omega)$, where $Y_2$ is the 
two-point cluster function as defined in Ref.~\cite{Mehta91}. \\

We will mainly analyze the correlations in the time domain, and by consequence
deal with the Fourier transforms of correlation functions. In general, we denote 
the Fourier transform of a given function of the energy $f(E)$ by:
\begin{equation}
\hat f(t) = {\cal F}[f](t) = \int\rmd\omega \; \rme^{2\pi\rmi\, \omega t} \;
f(\omega d) \; ,
\label{C_Ftrafo}\end{equation}
where the Fourier transform is taken with respect to the energy measured 
in units of $d$, and the factor $2\pi$ in the exponent assures proper 
normalization. For two spectral functions $f$ and $g$, the following relation
holds:
\begin{equation}
\hat C[f,g](t) = \frac{1}{L} \left\{ \lla \hat f'(-t) \; \hat g'(t) \rra -
\la \hat f'(-t)\ra \; \la \hat g'(t)\ra \right\} \; ,
\label{C_cthm}\end{equation}
where the functions $f'(x)$ and $g'(x)$ are equal to $f(x)$ and $g(x)$ inside
the interval $|x/ d|<L/2$ and zero outside, and the limits 
$N,L\to\infty \, , \; L/N\to 0$ are taken. Equation~(\ref{C_cthm}) is based on 
the convolution theorem \cite{Bri74} applied to the fluctuating parts of the 
spectral functions $f$ and $g$, where the convolution integral is expressed as 
a correlation function as in Eq.~(\ref{C_Cfg}) assuming stationarity. The 
limit $N,L\to\infty$ is necessary, to allow the correlation function to go to 
zero quickly enough, so that the Fourier integral of the correlation function 
is well defined. The limit $L/N\to 0$ serves to obtain stationarity in 
the interval where the correlation function is calculated. In particular, the 
average level (or resonance) density and the average S-matrix should not vary 
noticeably in this interval. Equation~(\ref{C_cthm}) is used in the numerical 
calculation of the correlation functions. It turned out, that 
$L= N/2\, , \; N=300$ already gives well converged results.

Note that we measure the energy in units of the mean level spacing $d$. As a
result, the argument of a correlation function is dimensionless, and so is
the argument of its Fourier transform. Nevertheless the latter is denoted by
$t$, as its significance is still time---though measured in units of $d^{-1}$.

\paragraph*{Total cross sections}
Consider the correlation function of two total cross sections 
$\sigma_\tot^{(a)}$ and $\sigma_\tot^{(b)}$ with possibly different entrance 
channels $a$ and $b$. As the total cross sections depend linearly on the 
respective diagonal S-matrix elements [see Eq.~(\ref{M_optthm})] the 
correlation function can be expressed as follows:
\begin{eqnarray}
C[\sigma_\tot^a,\sigma_\tot^b] &=& C[S_{aa} + S_{aa}^*,S_{bb} + S_{bb}^*]
\nonumber\\
&=& 2\, {\rm Re}\left( C[S_{aa},S_{bb}] + C[S_{aa},S_{bb}^*] \right) 
\nonumber\\
&=& 2\, {\rm Re}\, C[S_{aa},S_{bb}^*] \; .
\label{C_Csigtot}\end{eqnarray}
The correlation function of nonconjugated elements 
$C[S_{aa},S_{bb}]$ vanishes \cite{Mel85}. Relation~(\ref{C_Csigtot}) is 
essential, as it relates experimentally accessible quantities to analytical 
results \cite{VWZ85,Fyo97,Gor99}. For the Fourier transform of 
Eq.~(\ref{C_Csigtot}) we get:
\begin{eqnarray}
\hat C[\sigma_\tot^a,\sigma_\tot^b](t) &=&
2\, {\cal F}\, {\rm Re}\, C[S_{aa},S_{bb}^*](t) \nonumber\\
&=& \hat C[S_{aa},S_{bb}^*](t) + \hat C[S_{aa},S_{bb}^*](-t) \nonumber\\
&=& \hat C[S_{aa},S_{bb}^*](t) \; ,
\label{C_FCsigtot}\end{eqnarray}
where it is assumed that $t>0$. Then $\hat C[S_{aa},S_{bb}^*](-t)$ vanishes, 
because of its negative argument \cite{Har92} (see also Sec.~\ref{sF}). 
For the sake of brevity let us think of a correlation function and its Fourier
transform as a single object represented in the energy domain or in the time 
domain, respectively, and call it simply ``correlation function'' or 
``C-function'' in either case.

\paragraph*{Partial cross sections}
Partial cross sections are given by  
$\sigma_{ab}= |\delta_{ab} - S_{ab}|^2$. The theoretical treatment of 
correlation functions of partial cross sections is complicated by the fact that 
one has to average over a product of four S-matrix elements. The insertion of 
the S-matrix elements as given in Eq.~(\ref{M_Smat2}) leads to a double 
sum of resonance terms. As an exact analytical treatment seems to be 
impossible, we employ the diagonal approximation, which consists in retaining 
the diagonal terms of the double sum only. This is justified for sufficiently 
weak coupling and leads to:
\begin{equation}
\sigma_{ab} \approx \sigma_{ab}' = \sum_{j=1}^N 
\frac{\gamma_{ja} \gamma_{jb}}{(E-E_j)^2 + \Gamma_j^2/4} \; , \quad
\gamma_{ja} = \left|\tilde V_{ja}\right|^2 \; .
\label{C_sig_ab}\end{equation}
The {\it r.h.s.} can be written as the imaginary part of a function 
${\cal s}_{ab}(E)$, which has the same pole structure as the S-matrix:
\begin{eqnarray}
\sigma_{ab}'(E) &=& -2\, {\rm Im}{\cal s}_{ab}(E) \; , \nonumber\\
{\cal s}_{ab}(E) &=& \sum_{j=1}^N \frac{1}{\Gamma_j}\,
\frac{\gamma_{ja}\gamma_{jb}}{E-E_j+\rmi\Gamma_j/2} \; .
\label{C_cals}\end{eqnarray}
Due to the linear relation between the diagonal approximation
$\sigma_{ab}'$ and the spectral function ${\cal s}_{ab}$, we may again express
correlation functions of the former by corresponding correlation functions of 
the latter. In fact, we have:
\begin{equation}
C[\sigma_{ab},\sigma_{cd}](\omega) \approx 
C[\sigma_{ab}',\sigma_{cd}'](\omega) =
2{\rm Re}\, C[{\cal s}_{ab},{\cal s}_{cd}^*](\omega) \; .
\label{C_cals2}\end{equation}
It remains to calculate the correlation function of ${\cal s}_{ab}$ and 
${\cal s}_{cd}^*$. This will be done in the following section using the 
rescaled Breit-Wigner approximation. In order to calculate the Fourier 
transform of Eq.~(\ref{C_cals2}), we note that the pole structure of 
${\cal s}_{ab}(E)$ is the same as that of the true S-matrix, so that 
$\hat C[{\cal s}_{ab},{\cal s}_{cd}^*](-t)$ is again zero (for $t>0$). 
Therefore, we obtain:
\begin{equation}
\hat C[\sigma_{ab},\sigma_{cd}](t) \approx 
\hat C[\sigma_{ab}',\sigma_{cd}'](t) =
\hat C[{\cal s}_{ab},{\cal s}_{cd}^*](t) \; .
\label{C_FCsigab}\end{equation}

\section{Rescaled Breit-Wigner approximation\label{R}}

In this section we calculate the correlation function of two arbitrary 
S-matrix elements using the Breit-Wigner approximation, followed by a 
phenomenological rescaling procedure. To this end Eq.~(\ref{C_cthm}) is 
used, which means that we first calculate the Fourier transform of the
respective S-matrix elements (this can be done exactly), and then we average
over the resonance parameters 
$\{ E_j, \Gamma_j, \tilde V_{j1},\ldots,\tilde V_{jM}\}_{1\le j\le N}$. The
average over the real parts of the S-matrix poles $\{E_j\}$ can still be done
in a formally exact manner, but then we have to use the approximation mentioned
above, in order to finish the task. To obtain the correlation functions of 
partial cross sections, the same steps have to be done with the matrix elements 
${\cal s}_{ab}$ instead [see Eq.~(\ref{C_cals})].

\subsection{Formally exact treatment \label{sF}}

Here, we do all those steps of the calculation which are exact. We first
calculate the Fourier transform of one S-matrix element, using its pole 
expansion~(\ref{M_Smat2}). For $t>0$, we get:
\begin{eqnarray}
\hat S_{ab}(-t) &=& \frac{-\rmi}{d}\sum_{j=1}^N \tilde V_{ja}\tilde V_{jb} 
\int_{-L/2}^{L/2}\rmd\omega\; \frac{\rme^{-2\pi\rmi\, \omega t}}
{\omega - \frac{E_j-\rmi\Gamma_j/2}{d}} \nonumber\\
&=& \frac{-\rmi}{d}\sum_{j=1}^L \tilde V_{ja}\tilde V_{jb} \; 
\rme^{-2\pi\rmi\, E_j t/d}\; \rme^{-\pi\Gamma_j t/d} \; .
\end{eqnarray}
Here it was used, that the poles with real parts outside the integration region 
do not contribute to the Fourier transform, and for those inside, it is well 
justified to extend the integration up to infinity because $\Gamma_j/L \ll 1$. 

In the same way, we may obtain an analogous expression for $\hat S_{cd}(t)$.
The Fourier transforms of the average S-matrix elements are taken into account
in the numerical calculation only. Here by contrast, we notice that the average
S-matrix elements are almost constant in the integration interval, which means
that in the limit $N,L\to\infty$ their Fourier transforms will become 
$\delta$-functions situated at $t=0$. Thus they play no role in the current
calculation which is restricted to $t>0$. Note that we may define
$\hat C[S_{ab},S_{cd}^*](0) = \lim_{t\to 0} \hat C[S_{ab},S_{cd}^*](t)$.
Inserting the expressions for $\hat S_{ab}(-t)$ and $\hat S_{cd}^*(t)$ into
Eq.~(\ref{C_cthm}) and ignoring the Fourier transforms of the average 
S-matrix elements, we obtain:
\begin{eqnarray}
\hat C[S_{ab},S_{cd}^*](t) &=& \frac{1}{d^2 L} \sum_{j,k=1}^L
\tilde V_{ja}\tilde V_{jb} \; \tilde V_{kc}^*\tilde V_{kd}^* \;
\rme^{-2\pi\rmi\, (E_j-E_k) t/d}\; \rme^{-\pi(\Gamma_j+\Gamma_k)t/d}
\nonumber\\
&=& \frac{1}{d^2}\left\{ \lla 
\tilde V_{ja}\tilde V_{jb}\tilde V_{jc}^*\tilde V_{jd}^* \;
\rme^{-2\pi\Gamma_j t/d}\rra + (L-1) \lla 
\tilde V_{ja}\tilde V_{jb}\tilde V_{kc}^*\tilde V_{kd}^* \;
\rme^{-\pi(\Gamma_j+\Gamma_k)t/d}\; \rme^{-2\pi\rmi\, (E_j-E_k) t/d}
\rra \right\} \; . \nonumber\\
&\quad & 
\end{eqnarray}
As the ensemble average is invariant for any permutation of the resonance 
indices, the double sum can be evaluated. In the final expression the 
resonance indices $j\ne k$ are arbitrary. Note that 
$\hat C[S_{ab},S_{cd}^*](t)$ vanishes for $t<0$, because in this case both
Fourier transforms $\hat S_{ab}(-t)$ and $\hat S_{cd}^*(t)$ vanish, as 
can be easily seen by applying the residue theorem ({\it cf.} also \cite{Har92}). 

At last, we average formally over the real parts $\{E_j\}$ of the 
S-matrix poles. For fixed values of the partial amplitudes and the total
widths, the average over
$-L\;\exp[-2\pi\rmi(E_j-E_k)t/d]$ for $L\to\infty$ gives the two-point form 
factor \cite{Mehta91} of the random sequence $\{E_j\}$, which we denote by
$\tilde b_2(t)$. In general, $\tilde b_2(t)$ still depends on the parameters 
fixed.  In the weak coupling limit, however, the positions of the
resonances on the one hand, and the partial amplitudes and total widths on the
other hand, become statistically independent, so that $\tilde b_2(t)$ converges 
to the two-point form factor of the closed system. After all we may write:
\begin{equation}
\hat C[S_{ab},S_{cd}^*](t) = \frac{1}{d^2}\left\{ \lla 
\tilde V_{ja}\tilde V_{jb}\tilde V_{jc}^*\tilde V_{jd}^* \;
\rme^{-2\pi\Gamma_j t/d}\rra 
 - \lla \tilde V_{ja}\tilde V_{jb}\tilde V_{kc}^*\tilde V_{kd}^* \;
\rme^{-\pi(\Gamma_j+\Gamma_k)t/d}\; \tilde b_2(t) \rra \right\} \; .
\label{C_FCS}\end{equation}
This is so far an exact but rather formal result. However, it clearly shows
that the correlation function is no direct measure for spectral
correlations. The first term in Eq.~(\ref{C_FCS}), which may very well
dominate the correlation function, contains the parameters of only one single
resonance. Therefore, it cannot describe correlations between different 
resonances that are the ones we are really interested in. It is typically a 
monotonously decreasing function, where the decay is governed by the average 
width of the resonances. It is the second term in Eq.~(\ref{C_FCS}), which 
contains the parameters of two different resonances. It vanishes completely if 
there are no correlations between them. For our investigations it is important
to find situations, where the first term is relatively small, so that one may 
retrieve as much information as possible on the correlations between different 
resonances.

\subsection{Approximation \label{sA}}

In order to evaluate the remaining averages in Eq.~(\ref{C_FCS}) we have
to introduce some approximations. To this end, consider the weak coupling 
limit: $\forall a : \kappa_a \to 0$. Using first order perturbation theory in 
the expression for the S-matrix, Eqs.~(\ref{M_Smat}) and (\ref{M_Smat2}), 
we get the {\it pure} Breit-Wigner approximation for the S-matrix:
\begin{equation}
S_{ab}(E) \approx \delta_{ab} -\rmi\sum_{j=1}^N 
\frac{V_{ja}\;V_{jb}}{E-\eps_j+\rmi\sum_c V_{jc}^2/2}\; .
\end{equation}
This amounts to make the following replacements in the pole expansion of the
S-matrix~(\ref{M_Smat2}): 
\begin{equation}
\tilde V_{ja} \to V_{ja} \; , \quad
\Gamma_j \to \sum_{c=1}^M V_{ja}^2 \; , \quad
E_j\to \eps_j \; ,
\end{equation}
where $\eps_j$ are the eigenvalues of $H_\intern$. Hence, in order to obtain
the correlation functions in the Breit-Wigner approximation, we simply do the
same replacements in Eq.~(\ref{C_FCS}). Then the partial amplitudes become 
real uncorrelated Gaussian random variables, the total widths 
become simple functions of the partial amplitudes, and $\tilde b_2(t)$ becomes
the two-point form factor $b_2(t)$ of the spectrum of $H_\intern$:
\begin{equation}
\hat C[S_{ab},S_{cd}^*](t) \approx \frac{1}{d^2}\left\{ \lla
V_{ja}V_{jb}V_{jc}V_{jd}\; \rme^{-2\pi\sum_c V_{jc}^2\; t/d}\rra
-  \lla V_{ja}V_{jb}V_{kc}V_{kd}\; 
\rme^{-\pi\sum_c(V_{jc}^2+V_{kc}^2)\; t/d}\rra b_2(t) \right\} \; .
\end{equation}
The remaining Gaussian averages are relatively simple, so that in many 
cases the respective correlation function can be calculated in closed form. 
Note that the averages are different from zero, only if all partial amplitudes 
appear in even powers. 

Unfortunately, the pure Breit-Wigner approximation drifts quickly away from the 
exact result, as the coupling to the continuum increases. The following 
phenomenological procedure improves the approximation considerably. It consists 
in rescaling the variance of the partial amplitudes as follows:
\begin{equation}
\la V_{ja}^2\ra \to \frac{\la V_{ja}^2\ra}{(1+\kappa_a)^2} \; .
\label{C_rescal}\end{equation}
We call the result the ``rescaled Breit-Wigner'' approximation. As shown in
Ref.~\cite{mydiss}, it leads to partial fluctuating cross sections of the 
Hauser-Feshbach type \cite{HauFes52}, showing elastic enhancements of $2$ (GOE 
case) and $3$ (POE case) in agreement with earlier theoretical results 
\cite{AWM75,Mue87,Mue88}. The occurrence of Ericson fluctuations \cite{Eri66} 
with the correlation length $\Gamma_{\rm C}= d\sum_{a=1}^M T_a/(2\pi)$ is also 
correctly described. In Sec.~\ref{V} it is shown that in the time domain this 
approximation gives results for the correlation functions between S-matrix 
elements which become exact as $t\to 0$.

Note that the rescaled Breit-Wigner approximation can be applied in a wide 
range of situations, while it conserves the simplicity of the pure Breit-Wigner 
approximation. This makes it a valuable tool for the statistical description of 
complex scattering systems. This approximation can be justified to some 
extent with the following reasoning: It is well known, that the properties 
of a scattering system are determined by the Satchler transmission matrix
\cite{Sat63}, if the entire process is occurring on two 
different time scales, one associated with direct processes and one with 
long time processes also called compound processes in Nuclear physics.
In the absence of direct reactions or after an Engelbrecht-Weidenm\" uller
transformation \cite{EngWei73}, this implies the dependence on the 
transmission coefficients only.
They are directly related to the variances of the partial amplitudes as 
mentioned above. The rescaled approximation thus implies that we use the 
transmission coefficients of the system rather than the coupling 
constants. This can be viewed as a nonperturbative input, or as using 
phenomenological parameters.

\subsubsection*{Total cross sections}

According to Eq.~(\ref{C_FCsigtot}), the C-function between two total 
cross sections is equal to the C-function between the respective diagonal 
S-matrix elements. Hence $\hat C[\sigma_\tot^{(a)},\sigma_\tot^{(b)}](t)$ is 
given by the {\it r.h.s.} of Eq.~(\ref{C_FCS}) setting $b=a$ and 
$d=c=b$. In order to calculate this C-function in the 
rescaled Breit-Wigner approximation, we use the Eqs.~(\ref{M_kap}) and 
(\ref{M_Tau}) to express the ensemble averages as integrals over 
normalized squared amplitudes $g_1,\ldots,g_M$:
\begin{equation}
\hat C[\sigma_\tot^{(a)},\sigma_\tot^{(b)}](t) \approx T_a T_b \left\{
\lla g_a g_b \; \rme^{-G\; t}\rra
- \lla g_a\; \rme^{-G\; t/2} \rra \; \lla g_b\; \rme^{-G\; t/2} \rra \;
b_2(t) \right\} \; ,
\label{C_FCS-1}\end{equation}
with $G = \sum_{c=1}^M T_c \; g_c$. The normalized squared amplitudes $g_c$ are 
distributed as the random variables $V_{ia}^2/\la V_{ia}^2\ra$, {\it i.e.} they
are Porter-Thomas distributed \cite{PorTho56}:
\begin{equation}
p(g_c)= \frac{1}{\sqrt{2\pi g_c}} \; \rme^{-g_c/2}\; ,\quad c=1,\ldots, M\; .
\label{C_PT}\end{equation}
Now, the remaining averages in Eq.~(\ref{C_FCS-1}) can be calculated easily 
\cite{mydiss}. For the auto C-function of the total cross section we get:
\begin{equation}
\hat C[\sigma_\tot^{(a)}](t) \approx T_a^2 \left\{
3\; (1+2T_a t)^{-5/2} \prod_{c\ne a} (1+2T_c t)^{-1/2}
 - (1+T_a t)^{-3} \prod_{c\ne a} (1+T_c t)^{-1} \;
b_2(t) \right\} \; .
\label{C_FCS-1a}\end{equation}
Note, that this result is very similar to the result of Ref.~\cite{AlhFyo98} for 
the spectral autocorrelation function of the photodissociation cross section in 
weakly coupled chaotic systems (there the rescaling is used also, though 
without mentioning it; note that these authors obtain a rescaled 
formula for their specific case directly as an approximation to exact 
results \cite{FyoAlh98}). For the cross C-function of two different total cross 
sections, we get:
\begin{eqnarray}
\hat C[\sigma_\tot^{(a)},\sigma_\tot^{(b)}](t) &\approx & T_a T_b 
 \left\{
(1+2T_a t)^{-3/2}(1+2T_b t)^{-3/2} \prod_{c\ne a,b} (1+2T_c t)^{-1/2} \right. 
\nonumber\\
&&\qquad\qquad\qquad \left. 
- (1+T_a t)^{-2} (1+T_b t)^{-2} \prod_{c\ne a,b} (1+T_c t)^{-1}
\; b_2(t) \right\} \; .
\label{C_FSC-1b}\end{eqnarray}

In what follows, we assume that all coupling strengths are equal: 
$\forall a : \kappa= \kappa_a, \, T= T_a$. For convenience, the numerical 
analysis in Sec.~\ref{N} is restricted to this case only. The 
Eqs.~(\ref{C_FCS-1a}) and (\ref{C_FSC-1b}) simplify considerably and can 
be combined into a single one:
\begin{equation}
\hat C[\sigma_\tot^{(a)} , \sigma_\tot^{(b)}] \approx T^2 
\left\{ (1+2\delta_{ab}) \; (1+2T t)^{-2-M/2}
 - (1+T t)^{-2-M} \; b_2(t) \right\} .
\label{C_FCS-2}\end{equation}
To lowest order in the transmission coefficient $T$ the {\it r.h.s.} becomes 
$T^2 \, [1+2\delta_{ab}-b_2(t)]$ in agreement 
with results on intensity weighted stick spectra \cite{Lom93,AlGuHa97} and
the asymptotic behavior of the exact analytical result for the GOE case
(see Eq.~(\ref{V_FVWZaT2}) in Sec.~\ref{V}). The difference in the 
C-function at small times between GOE and POE is known as the correlation 
hole. In order to quantify it, we use its size at $t=0$ relative to the maximal 
size of the C-function in the POE case. From Eq.~(\ref{C_FCS-2}) it
follows that the correlation hole is $1/3$ in the case of the auto C-function 
($a=b$), while it is one in the case of the cross C-function ($a\ne b$).

\subsubsection*{Partial cross sections}

According to Eq.~(\ref{C_FCsigab}), the C-function of two partial cross 
sections in the diagonal approximation is equal to the C-function of the 
matrix elements ${\cal s}_{ab}$ and ${\cal s}_{cd}^*$, defined in 
Eq.~(\ref{C_cals}). Following the same lines as in the case of the true
S-matrix, we arrive at the following expression for the latter C-function:
\begin{equation}
\hat C[{\cal s}_{ab},{\cal s}_{cd}^*](t) \approx \frac{4\pi^2}{d^2} \left\{
\lla \frac{\gamma_{1a} \gamma_{1b} \gamma_{1c} \gamma_{1d}}{\Gamma_1^2} \;
\rme^{-2\pi\;\Gamma_1\; t/d} \rra
 - \lla \frac{\gamma_{1a}\gamma_{1b}}{\Gamma_1}
\; \frac{\gamma_{2c}\gamma_{2d}}{\Gamma_2} \; 
\rme^{-\pi\;(\Gamma_1+\Gamma_2) \; t/d} \; \tilde b_2(t) \rra \right\} \; ,
\label{C_FCSpart-1}\end{equation}
which is the analog of Eq.~(\ref{C_FCS}). To this expression we can 
apply the rescaled Breit-Wigner approximation. The partial 
widths $\gamma_{ja}$, are replaced by Porter-Thomas distributed random 
variables with average value $\la\gamma_{ja}\ra = T_a\, d/(2\pi)$, the
total widths become the sums of the partial widths, and the two-point
form factor $\tilde b_2(t)$ becomes the two-point form factor of the spectrum
of $H_\intern$. Thus we get:
\begin{equation}
C[\sigma_{ab},\sigma_{cd}](t) \approx T_a T_b T_c T_d \left\{ 
\lla \frac{g_a\, g_b\, g_c\, g_d}{G^2} \;
\rme^{-G\, t}\rra
 - \lla \frac{g_a\, g_b}{G} \; \rme^{- G\, t/2}\rra
\lla\frac{g_c\, g_d}{G} \; \rme^{- G\, t/2} \rra \; b_2(t) \right\} \; ,
\label{C_FCSpart-2}\end{equation}
where $G= \sum_c T_c g_c$. The ensemble average can be performed, using the 
following identities:
\begin{eqnarray}
G^{-1}\;\rme^{-\alpha G} &=& \int_\alpha^\infty\rmd\alpha' \;\rme^{-\alpha' G}
\; , \nonumber\\
G^{-2}\;\rme^{-\alpha G} &=& \int_\alpha^\infty\rmd\alpha'
\int_{\alpha'}^\infty\rmd\alpha''\; \rme^{-\alpha'' G} \; ,\quad 
\alpha > 0 \; .
\label{C_FCStrick}\end{eqnarray}
Exchanging the integration on $\alpha'$ and $\alpha''$ with the ensemble
average, we can do the ensemble average analytically, and we are only left 
with the integrals over the auxiliary variables $\alpha'$ and $\alpha''$. With 
the exception of some special cases, those integrals have to be evaluated 
numerically (for details, see Ref.~\cite{mydiss}). In the case of equal coupling
strengths, however, we obtain the following analytical expression:
\begin{equation}
\hat C[\sigma_{ab},\sigma_{cd}](t) \approx \frac{T^2}{(2+M)^2} \left\{
\frac{(1+2/M)^2}{(1+6/M)\, (1+4/M)} \; A \;
 (1+2Tt)^{-2-M/2} - B\; (1+Tt)^{-2-M} \, b_2(t) \right\} \, .
\label{R_FCsigab}\end{equation}
Here, $A$ and $B$ depend on the actual combinations of partial widths
in Eq.~(\ref{C_FCSpart-2}), {\it i.e.} whether any of the channels 
involved coincide or not. The following cases occur in the present paper:
\begin{itemize}
\item{The autocorrelation function: \\
$\hat C[\sigma_{ab}]\, :\; A= 9+96\, \delta_{ab}\, ,\; 
B= 1+8\, \delta_{ab}\, .$}
\item{The cross correlation function between two different elastic cross 
sections: \\
$\hat C[\sigma_{aa},\sigma_{bb}]\, :\; A=9\, ,\; B=9\, .$}
\item{The cross correlation function between an elastic and an inelastic
cross section:\\
$\hat C[\sigma_{aa},\sigma_{ab}]\, :\; A=15\, ,\; B=3\, .$}
\end{itemize}
According to Eq.~(\ref{R_FCsigab}), the coefficients $A$ and $B$ determine
the relative depth of the correlation hole in the limit of many channels 
$M\to\infty$. The prefactor $(1+2/M)^2/[(1+6/M)(1+4/M)]$ in front of $A$ may,
however, lead to a considerably deeper correlation hole, as long as $M$ is not
too large.

For $M\to\infty$ and $MT$ fixed, the algebraic decay in 
Eq.~(\ref{R_FCsigab}) turns into an exponential one: 
\begin{equation}
(1+2Tt)^{-2-M/2} \sim (1+Tt)^{-2-M} \sim \exp(-MT\, t) \; .
\end{equation}
This means, that we obtain Ericson fluctuations \cite{Eri66} in this limit. 
The reason for the occurrence of Ericson fluctuations can be
understood from Eq.~(\ref{C_FCSpart-2}): Only in the case of many 
channels and small transmission coefficients, does the central limit theorem 
lead to negligible fluctuations of the total width around its average value. 
Then we may treat $G$ as a constant, which leads immediately to the expected
exponential decay. Note that this implies that Ericson fluctuations are no 
reliable signature for chaotic scattering, because the central limit theorem 
may work, even if the partial amplitudes are not Gaussian distributed, thus 
leading again to Ericson fluctuations. In such a situation, the deviation from 
the exponential decay due to $b_2(t)$, would be the only reliable signature of 
the chaotic dynamics.

\section{\label{V}The Fourier transform of the VWZ integral}

In this section we derive a general formula for the Fourier transform of the 
VWZ integral \cite{VWZ85}. This allows to obtain exact results for the
correlation functions of total cross sections in the GOE case. We need these 
results, in order to check the accuracy of the rescaled Breit-Wigner 
approximation. Apart from that it will turn out, that numerically it is much
easier to calculate the Fourier transform than the original VWZ integral:
\begin{eqnarray}
C[S_{ab},S_{cd}^*](\omega) &=& \frac{1}{8} \int\int_0^{\;\infty}
\rmd\lambda_1 \, \rmd\lambda_2 \int_0^1\rmd\lambda \;
\frac{\lambda (1-\lambda) |\lambda_1-\lambda_2|
\; \rme^{-\rmi\pi\omega\, (\lambda_1+\lambda_2+2\lambda)}}
{\sqrt{\lambda_1 (1+\lambda_1) \; \lambda_2 (1+\lambda_2)} \;
(\lambda+\lambda_1)^2 (\lambda+\lambda_2)^2} \nonumber\\
&&\qquad\times \; \prod_{e=1}^M
\frac{1- T_e \lambda}{\sqrt{(1+T_e \lambda_1)(1+T_e \lambda_2)}} \;
\left\{ \delta_{ab}\delta_{cd} \; \Delta_a \; \Delta_c \; + \;
(\delta_{ac}\delta_{bd} + \delta_{ad}\delta_{bc}) \;\Pi_{ab} \right\} \; ,
\label{V_VWZint}\\
\Delta_a &=& T_a \sqrt{1-T_a} \; \left( \frac{\lambda_1}{1+T_a \lambda_1} +
\frac{\lambda_2}{1+T_a \lambda_2} + \frac{2\lambda}{1-T_a \lambda} \right)
\; , \nonumber\\
\Pi_{ab} &=& T_a T_b \; \left(
\frac{\lambda_1\; (1+\lambda_1)}{(1+T_a \lambda_1)\; (1+T_b \lambda_1)} +
\frac{\lambda_2\; (1+\lambda_2)}{(1+T_a \lambda_2)\; (1+T_b \lambda_2)} +
\frac{2\lambda \; (1-\lambda)}{(1-T_a \lambda)\; (1-T_b \lambda)}
\right) \; . \nonumber
\end{eqnarray}
The main interest in the VWZ integral has been the calculation of average 
fluctuating cross sections \cite{Mel85,Ver86,HarHue87,HarHue90},
which corresponds to the case $\omega =0$ in 
Eq.~(\ref{V_VWZint}). Only a few papers treat the $\omega$-dependence of 
the VWZ integral \cite{Har92,MueHar90,Dit01}, and even then the 
analysis was usually restricted to particular limits such as $\omega\to 0$  
($t\to\infty$ in the time domain), or many channels: $M\to\infty$ 
and Ericson fluctuations. Quite often the existence of the correlation hole 
was simply ignored.

In the present analysis, the correlation hole and its dependence on the 
coupling strengths is of great importance. It is needed to distinct regular 
from chaotic dynamics. Therefore, we will analyze the Fourier transform of the
VWZ integral in some detail. We will also observe the behavior of the
correlation hole in the limits: $t\to 0$ (Sec.~\ref{VST}) and 
$\forall a : T_a\to 0$ (Sec.~\ref{VSC}). \\

Let us start our derivation by applying the Fourier transform (\ref{C_Ftrafo}) 
to Eq.~(\ref{V_VWZint}). We exchange the Fourier integration on $\omega$ 
with the integrals on $\lambda,\lambda_1$, and $\lambda_2$. For the Fourier 
integration we then simply need to calculate:
\begin{eqnarray}
\int\rmd\omega \; \exp\left\{ 2\pi\rmi\omega\, \left[\, t -
(\lambda_1 +\lambda_2 + 2\lambda)/2\right] \right\}  \nonumber\\
= \delta\left[\, t - (\lambda_1 +\lambda_2 + 2\lambda)/2\right] \; .
\end{eqnarray}
The $\delta$-function can be used to remove the $\lambda$-integral. To the
remaining double integral, the following transformations are applied:
\begin{eqnarray}
\lambda_1\, ,\,\lambda_2 \; \to \; 
r= \frac{\lambda_1 +\lambda_2}{2} \; , \quad s= \lambda_2 - \lambda_1 
\nonumber\\
\mbox{followed by} \quad s \; \to \;  x = r^2 - s^2/4  \; .
\end{eqnarray}
This leads to:
\begin{equation}
\hat C[S_{ab},S_{cd}^*](t) = \frac{1}{4}\int_{\max(0,t-1)}^t \rmd r \;
(t-r) (r+1-t) \prod_{e=1}^M\left[1-T_e(t-r)\right] \; U(r) \; ,
\label{V_FVWZ}\end{equation}
where
\begin{eqnarray}
U(r) &=& 2\int_0^{r^2}\rmd x \; 
\frac{\delta_{ab}\delta_{cd}\; \Delta_a \Delta_c
\; + \; (\delta_{ac}\delta_{bd}+\delta_{ad}\delta_{bc}) \; \Pi_{ab}}
{(t^2-r^2+x)^2 \; \sqrt{x(x+2r+1)} \; \sqrt{\prod_{e=1}^M(1+2T_e r+T_e^2 x)}}
\; , \label{V_U1}\\
\Delta_a &=& 2T_a \sqrt{1-T_a} \left( \frac{r+T_a x}{1+T_a(2r+T_a x)} +
\frac{t-r}{1-T_a(t-r)} \right) \; ,\, \mbox{and}\label{V_Delta}\\
\Pi_{ab} &=& 2 T_a T_b \left(
\frac{T_a T_b x^2 + [T_a T_b r + (T_a+T_b)(r+1)-1] x + r(2r+1)}
{(1+2T_a r+T_a^2 x)(1+2T_b r+T_b^2 x)}
+ \frac{(t-r)(r+1-t)}{[1-T_a(t-r)] \; [1-T_b(t-r)]} \right) \; . \nonumber\\
 &\quad &
\label{V_Pi}\end{eqnarray}
In order to remove the $1/\sqrt{x}$-singularity in the integrand of $U(r)$, we 
substitute consecutively:
\begin{equation}
x= (y-1)b/2 \; , \quad y= (z+z^{-1})/2 \; , \quad z= 2u +1 \quad
\mbox{where}\quad b= 1+2r \; .
\end{equation}
This gives:
\begin{equation}
U(r) = 4\int_0^r\frac{\rmd u}{2u+1} \; 
\frac{\delta_{ab}\delta_{cd}\; \Delta_a \Delta_c \; +\; 
(\delta_{ac}\delta_{bd}+\delta_{ad}\delta_{bc}) \; \Pi_{ab}}
{(t^2-r^2+x)^2 \; \sqrt{\prod_{e=1}^M(1+2T_e r+T_e^2 x)}} \; , \quad
x= \frac{bu^2}{2u+1} \; .
\label{V_U2}\end{equation}
The Eqs.~(\ref{V_FVWZ}) and (\ref{V_U2}) together with the Eqs.~(\ref{V_Delta})
and (\ref{V_Pi}) form our final result for the correlation function 
$\hat C[S_{ab},S_{cd}^*](t)$ (it is understood, to replace $x$ by $bu^2/(2u+1)$ 
wherever it occurs). One may readily use these formulas for numerical 
calculations. All problematic singularities have been removed from the 
integration region. Our result if followed by a fast Fourier transformation
back to the energy domain, may even be a quite efficient way of computing
the original VWZ integral. \\

\begin{figure}
\input{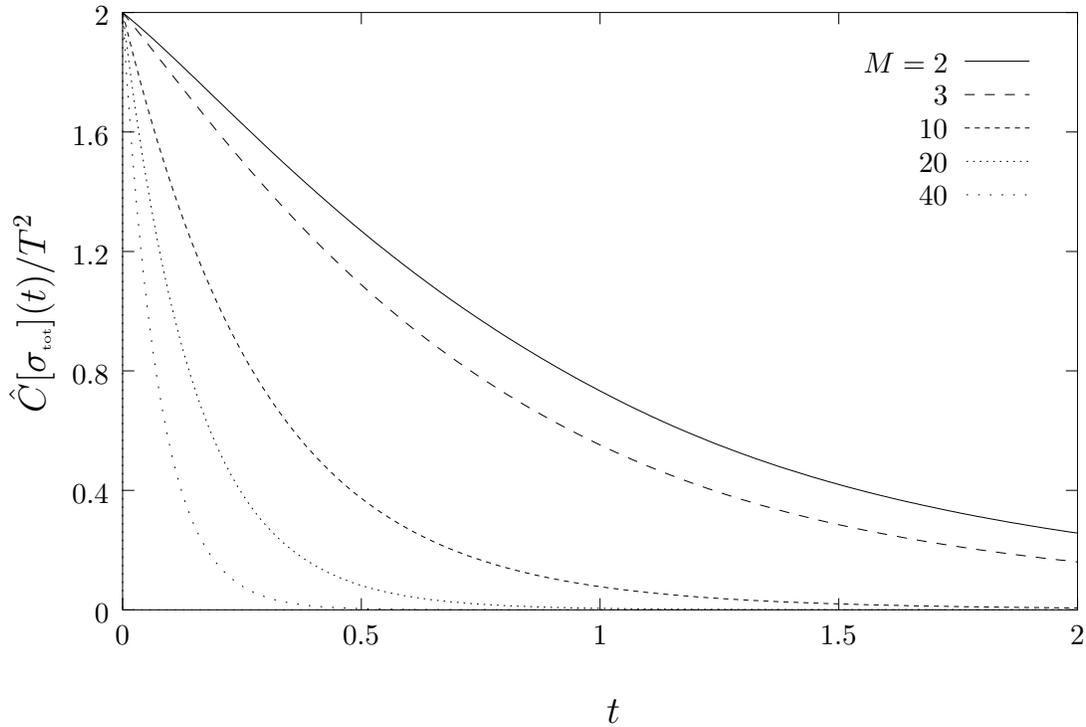}
\caption{The exact result (\ref{V_FVWZ}) for the autocorrelation function 
$\hat C[\sigma_\tot](t)$ divided by $T^2$. While $\kappa = 0.1$ is kept fixed,
the number of channels is varied: $M=2$ (solid line), $M=3$ (long dashed line),
$M=10$ (short dashed line), $M=20$ (dotted line), and $M=40$ (long dotted 
line).}
\label{f_C1-1auto}\end{figure}

\begin{figure}
\input{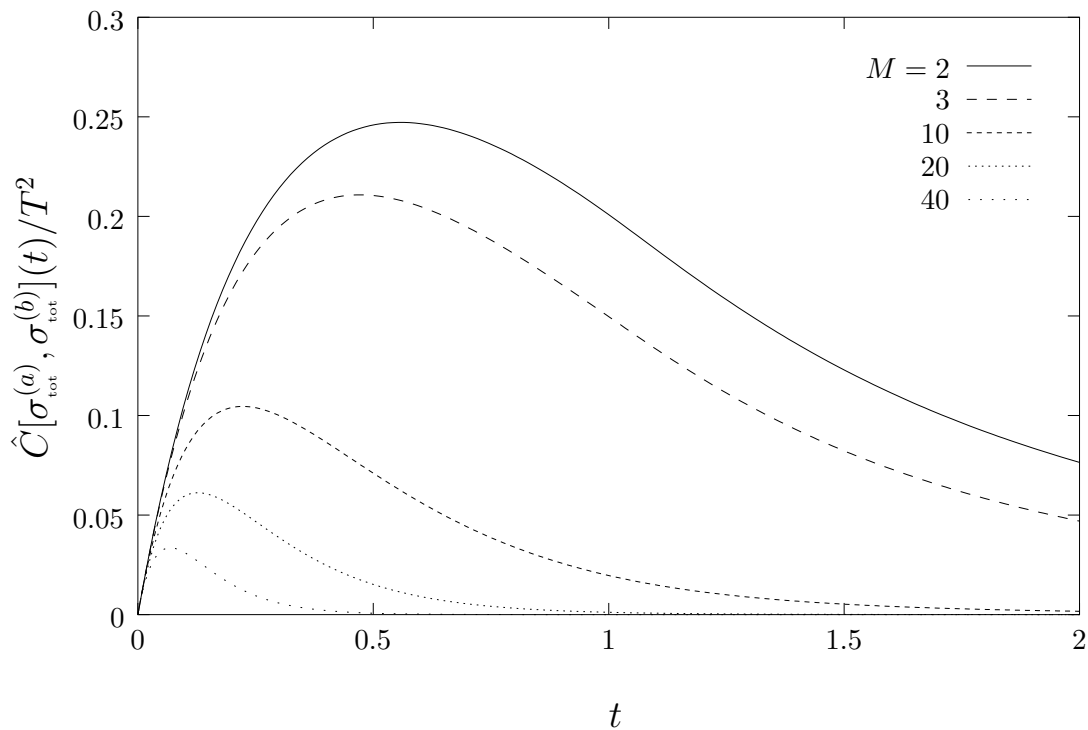}
\caption{As in Fig.~\ref{f_C1-1auto}, but for the cross correlation 
function $\hat C[\sigma_\tot^a,\sigma_\tot^b](t)$ divided by $T^2$.} 
\label{f_C1-1cross}\end{figure}

The Figs.~\ref{f_C1-1auto} and \ref{f_C1-1cross} show examples of the 
Fourier-transformed auto and cross correlation functions of the total cross 
section, calculated with the help of Eq.~(\ref{V_FVWZ}), etc. The coupling 
strength for all channels was fixed to the value $\kappa= 0.1$, while the 
number of channels $M$ was varied. The qualitative features of the C-functions
are almost the same for any $M$. In the first place, $M$ determines the 
falloff at large times. Note, however, that in the case of extremely small 
total transmission, we would obtain a ``true'' correlation hole at $t=0$ 
({\it i.e.} a positive slope). See also the related discussion below 
Eq.~(\ref{C_FCS-2})] which is concerned with the rescaled Breit-Wigner
result. In all, it permits us to restrict the numerical studies 
in Sec.~\ref{N} to the case $M=2$. 

Figure~\ref{f_C1-1auto} shows the autocorrelation function divided by $T^2$,
which starts at $t=0$ with the value $2$, independent of the number of 
channels. It decays with time, the larger the number of channels, the faster. 
This is very reasonable, as the decay must be related to the total transmission 
$\sum_a T_a = M T$. Note that there is no qualitative signature (positive slope
at $t=0$) of the correlation hole.

In Fig.~\ref{f_C1-1cross} the cross correlation function is shown, which
starts at $t=0$ with zero. It increases linearly with time, and it seems that 
now the slope is independent of the number of channels. At large times, the 
cross C-function must go to zero in the same way as the auto C-function shown 
in Fig.~\ref{f_C1-1auto}. Hence, after reaching a maximum that decreases and 
moves to the left with increasing $M$, the cross C-function decays to zero. 
Therefore, we see a clear signature of the correlation hole. \\

To conclude this section we will consider two different asymptotic limits: 
First we will prove, that for $t\to 0$ the asymptotic behavior of the 
various C-functions is independent of $M$. Second we will show with very high
numerical precision, that for small transmission coefficients, 
$\forall a\, :\, T_a\to 0$, the cross C-functions coincide (if properly 
normalized) with the two-level form factor $b_2(t)$ for the GOE.

\subsection{\label{VST}The limit of small times}

In the limit $t\to 0$, the Eq.~(\ref{V_FVWZ}) can be approximated by:
\begin{eqnarray}
\hat C[S_{ab},S_{cd}^*](t) &\sim&
\frac{t^2}{4}\int_0^1\rmd\varrho \; (1-\varrho) \; U(t\varrho) \nonumber\\
&=& \frac{1}{4}\int_0^1\rmd\varrho \; (1-\varrho) \; \tilde U(\varrho) \; ,
\end{eqnarray}
where $\tilde U(\varrho) = t^2 U(t\rho)$. Hence setting $r= t\rho$ in
Eq.~(\ref{V_U2}) the following limits can by taken: $b\sim 1$, 
$2u+1 \sim 1$, and $x \sim u^2$. Then substituting $u= tv$ we get:
\begin{eqnarray}
&&\tilde U(\varrho) \sim \frac{4}{t}\int_0^1\rmd v \left[
\delta_{ab}\delta_{cd}\; \Delta_a \Delta_c \; +\; 
(\delta_{ac}\delta_{bd}+\delta_{ad}\delta_{bc}) \; \Pi_{ab}\right]
\nonumber\\
&&\qquad\qquad \div \left[\textstyle
(1-\varrho^2+v^2)^2 \sqrt{\prod_{e=1}^M(1+2T_e t\varrho + T_e^2 t^2 v^2)}
\right] \nonumber\\
&&\quad \sim \frac{4}{t}\int_0^1\rmd v\; \frac{
\delta_{ab}\delta_{cd}\, \Delta_a \Delta_c +
(\delta_{ac}\delta_{bd}+\delta_{ad}\delta_{bc}) \, \Pi_{ab}}
{(1-\varrho^2+v^2)^2} \; .
\end{eqnarray}
To lowest order the Eqs.~(\ref{V_Delta}) and (\ref{V_Pi}) simplify also:
\begin{eqnarray}
\Delta_a &\sim& 2T_a \sqrt{1-T_a} \; t \; (1+T_a t v^2) 
\sim 2T_a \sqrt{1-T_a} \; t \\
\Pi_{ab} &\sim& 2 T_a T_b \; t \; \left[ 1+ (T_a + T_b-1) t v^2\right] 
\sim 2 T_a T_b \; t \; .
\end{eqnarray}
This shows, that if the C-function contains the $\Delta$-term only, we get
a linear increase starting at $t=0$, whereas if it contains the $\Pi$-term 
also, we get a finite value at $t=0$, but no valid value for the slope. The 
slope in the first case and the finite value in the second are both determined 
by the same integral $I_1$:
\begin{eqnarray}
\hat C[S_{ab},S_{cd}^*](t) &\sim & \left[ 4\; \delta_{ab}\delta_{cd} \; T_a T_c \;
\sqrt{(1-T_a)(1-T_c)}\; t + 2\; (\delta_{ac}\delta_{bd}+\delta_{ad}\delta_{bc}) \;
T_a T_b \right] I_1 \; , \\
I_1 &=& \int_0^1\rmd\varrho \; (1-\varrho) \;
\int_0^1\frac{\rmd v}{(1-\varrho^2+v^2)^2}
= \int_0^1\rmd\varrho \; (1-\varrho) \; \left[
\frac{1}{2p^3}\arctan\left(\frac{\varrho}{p}\right) + \frac{\varrho}{2p^2}
\right] \; ,
\end{eqnarray}
where $p^2 = 1- \varrho^2$. Splitting $I_1$ into two parts and substituting 
$x= \arctan(\varrho/p)$ in the first one, we can evaluate $I_1$ analytically:
\begin{equation}
I_1 = \frac{1}{2} \int_0^{\pi/2}\rmd x \; \frac{x}{1+\sin x} + 
\frac{1}{2} \int_0^1\rmd\varrho \; \frac{\varrho}{1+\varrho}
= \frac{\ln 2}{2} + \frac{1-\ln 2}{2} = \frac{1}{2} \; .
\end{equation}
In order to contain the $\Delta$-term only, the C-function must be of the 
form: $\hat C[S_{aa},S_{bb}^*]$ with $a\ne b$. Then its asymptotic behavior
is:
\begin{equation}
\hat C[S_{aa},S_{bb}^*] \sim 2 \; T_a T_b \; \sqrt{(1-T_a)(1-T_b)} \; t \; .
\label{V_FDelta}\end{equation}
If the C-function is of the form 
$\hat C[S_{ab},S_{ab}^*] = \hat C[S_{ab},S_{ba}^*]$, then it contains the 
$\Pi$-term also. In this case we get:
\begin{equation}
\hat C[S_{ab},S_{ab}^*](0) = (1+\delta_{ab}) \; T_a T_b \; .
\label{V_FPi}\end{equation}
These results confirm our observations in Fig.\ \ref{f_C1-1auto} and 
Fig.\ \ref{f_C1-1cross}. The cross correlation function of two total cross
sections vanishes at $t=0$. This holds for all values of the transmission 
coefficient. Moreover, the slope only depends on the transmission coefficients 
of the entrance and exit channels, and thus does not change if the number of 
channels is changed.  In contrast to that, for the auto correlation function of 
a total cross section we get: $\hat C[\sigma_\tot^{(a)}](0) = 2 T_a^2$. Again
this value does not depend on the number of channels.

Comparing the asymptotic results (\ref{V_FDelta}) and (\ref{V_FPi}) to the
rescaled Breit-Wigner approximation (\ref{C_FCS-2}) we find agreement in lowest 
order---for the slope in the case of cross correlation functions, and for the 
value at $t=0$ in the case of autocorrelation functions.

\subsection{\label{VSC}The limit of small transmission coefficients}

In the case $\forall a\, :\, T_a\to 0$, the Eqs.~(\ref{V_FVWZ}) and
(\ref{V_U2}) simplify to:
\begin{equation}
\hat C[S_{ab},S_{cd}^*](t) \sim \frac{1}{4}\int_{\max(0,t-1)}^t \rmd r \;
(t-r) (r+1-t) \; U(r)  \; ,
\label{V_FVWZaT}\end{equation}
where
\begin{equation}
U(r) \sim 4 \int_0^r \frac{\rmd u}{2u+1} \; 
\frac{\delta_{ab}\delta_{cd}\, \Delta_a \Delta_c +
(\delta_{ac}\delta_{bd}+\delta_{ad}\delta_{bc}) \, \Pi_{ab}}
{(t^2-r^2+ \frac{bu^2}{2u+1})^2} \; .
\end{equation}
In the same limit, the expression for $\Delta_a$ (\ref{V_Delta}) and the one for
$\Pi_{ab}$ (\ref{V_Pi}) become:
\begin{eqnarray}
\Delta_a &\sim& 2 T_a \; t \\
\Pi_{ab} &\sim& 2 T_a T_b
\left[ (t-r)(r+1-t) + r(2r+1) - \frac{bu^2}{2u+1} \right] \; .
\end{eqnarray}
Defining the two integrals:
\begin{eqnarray}
I_2 &=& \int_0^r\rmd u \; \frac{2u+1}{\left[ (t^2-r^2)(2u+1) + bu^2\right]^2}
\\
I_3 &=& \int_0^r\rmd u \; \frac{u^2}{\left[ (t^2-r^2)(2u+1) + bu^2\right]^2}
\; ,
\end{eqnarray}
the asymptotic limit of $U(r)$ can be written as
\begin{equation}
U(r) \sim 8 \left\{ 2 \delta_{ab}\delta_{cd}\; T_a T_c \; t^2 \; I_2 
+ (\delta_{ac}\delta_{bd}+\delta_{ad}\delta_{bc}) \;
T_a T_b \left[ (r^2+2tr+t-t^2) I_2 - b I_3\right] \right\} \; .
\label{V_UaT}\end{equation}
In order to keep the discussion short, we shall consider cross C-functions of the
type $\hat C[S_{aa},S_{bb}^*] ,\, a\ne b$ only. Then we only need to calculate 
$I_2$. Defining $p^2= t^2-r^2$ and $R^2= b-p^2$, $I_2$ can be written as follows:
\begin{eqnarray}
I_2 &=& -\frac{1}{2p} \partial_p \int_0^r\frac{\rmd u}{bu^2+2p^2 u+p^2} 
\nonumber\\
&=& -\frac{1}{2p} \partial_p \; \frac{1}{pR}\left(
\arctan\frac{br+p^2}{pR} - \arctan\frac{p}{R}\right) \nonumber\\
&=& \frac{R^2-p^2}{2p^3 R^3}\left( \arctan\frac{br+p^2}{pR} -
\arctan\frac{p}{R}\right) - \frac{r(r+1)}{2t^2 p^2 R^2} \nonumber\\
&=& \frac{R^2-p^2}{2p^3 R^3} \arctan\frac{rR}{(r+1)p} + 
\frac{r(r+1)}{2t^2 p^2 R^2} \; .
\label{V_I2}\end{eqnarray}
Inserting Eq.~(\ref{V_UaT}) into Eq.~(\ref{V_FVWZaT}), we get the following 
result for $\hat C[S_{aa},S_{bb}^*]$:
\begin{equation}
\frac{1}{T_a T_b} \; \hat C[S_{aa},S_{bb}^*](t) 
\sim 4 t^2 \int_{\max(0,t-1)}^t\rmd r\; (t-r) (r+1-t) \; I_2 \; ,
\label{V_FVWZaT2}\end{equation}
where $I_2$ is given in Eq.~(\ref{V_I2}). We have evaluated Eq.~(\ref{V_FVWZaT2}) 
numerically and compared to the expected result $1-b_2(t)$, for the discrete 
GOE-spectrum. The difference is of the order of the machine precision 
($\approx 10^{-11}$) for any value of $t$.

\section{\label{T}Test of the rescaled Breit-Wigner approximation}

The aim of the present paper is actually twofold. In the first place, we
analyze correlation functions of total and partial cross sections, to search 
for signatures of the correlation hole, in particular in the regime of 
overlapping resonances. In the second place we wish to establish the rescaled 
Breit-Wigner approximation as a simple and general tool for the analysis of 
correlations in different scattering situations. This section is devoted to 
the latter, whereas the the correlation hole will be studied in 
Sec.~\ref{N}.

We use the rescaled Breit-Wigner approximation in those cases, where no
exact theory is available, {\it i.e.} for C-functions of partial cross sections
(where we need the additional diagonal approximation) and for C-functions of 
total cross sections, in the POE case. In this section we use the C-functions
of total cross sections in the GOE case (for which we have exact analytic 
results) to check the validity of the rescaled Breit-Wigner approximation. \\

\begin{figure}
\input{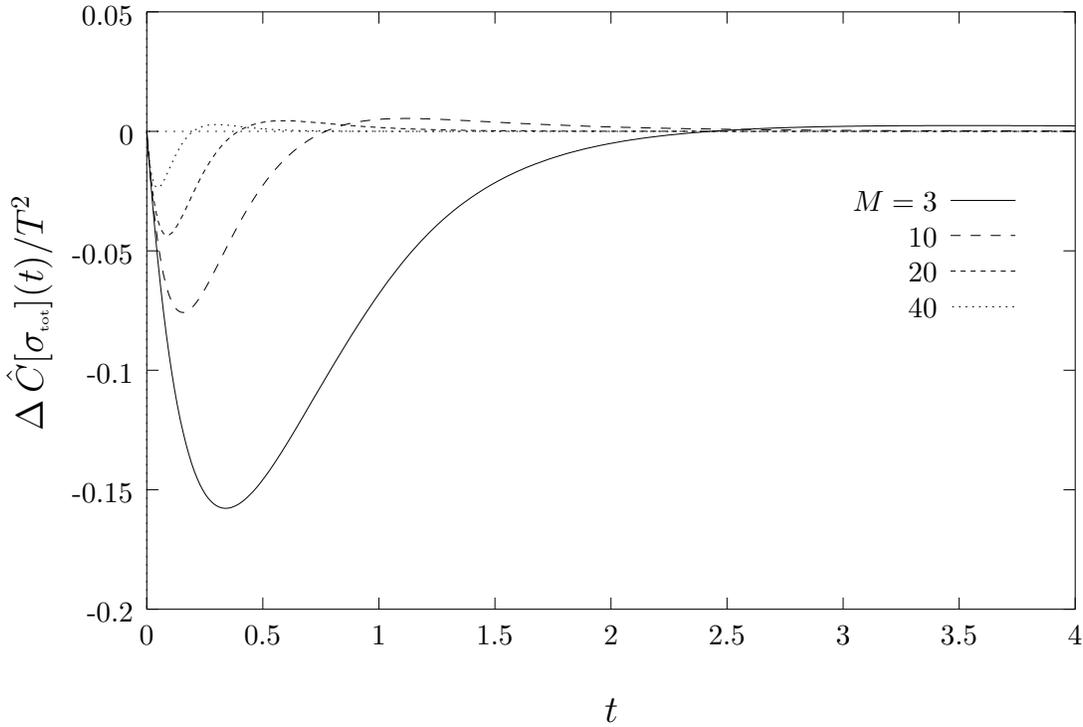}
\caption{Difference between the rescaled Breit-Wigner approximation
(\ref{C_FCS-2}) and the exact result (\ref{V_FVWZ}) for the autocorrelation 
function $\hat C[\sigma_\tot](t)$ divided by $T^2$. The coupling
strength is kept fixed $\kappa = 0.1$, and the number of channels is varied:
$M=3$ (solid line), $M=10$ (long dashed line), $M=20$ (short dashed line), and 
$M=40$ (dotted line).}
\label{f_d1-1auto}\end{figure}

\begin{figure}
\input{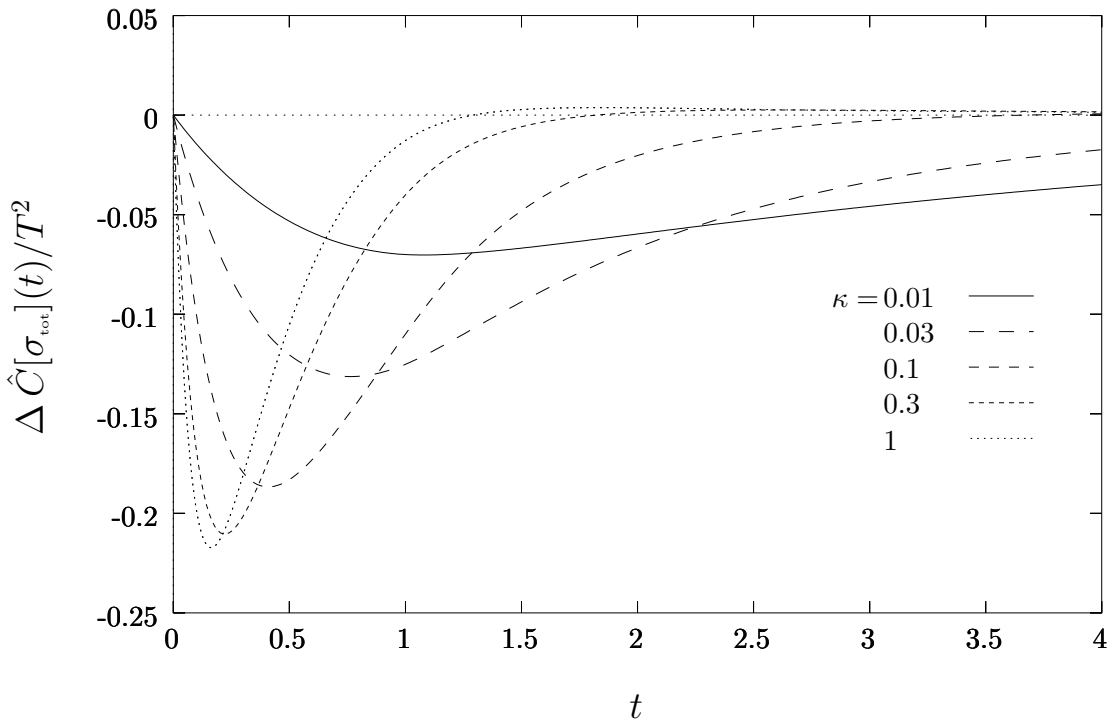}
\caption{As in Fig.~\ref{f_d1-1auto}, but here, the number of channels $M=2$ is 
kept fixed, and the coupling strength is varied: $\kappa = 0.01$ (solid line), 
$\kappa =0.03$ (long dashed line), $\kappa = 0.1$ (medium dashed line), 
$\kappa = 0.3$ (short dashed line), and $\kappa = 1$ (dotted line).}
\label{f_dK2auto}\end{figure}

\begin{figure}
\input{F05_d1-1cross.pstex_t}
\caption{As in Fig.~\ref{f_d1-1auto}, but for the cross correlation function 
$\hat C[\sigma_\tot^a,\sigma_\tot^b](t)$ divided by $T^2$.}
\label{f_d1-1cross}\end{figure}

\begin{figure}
\input{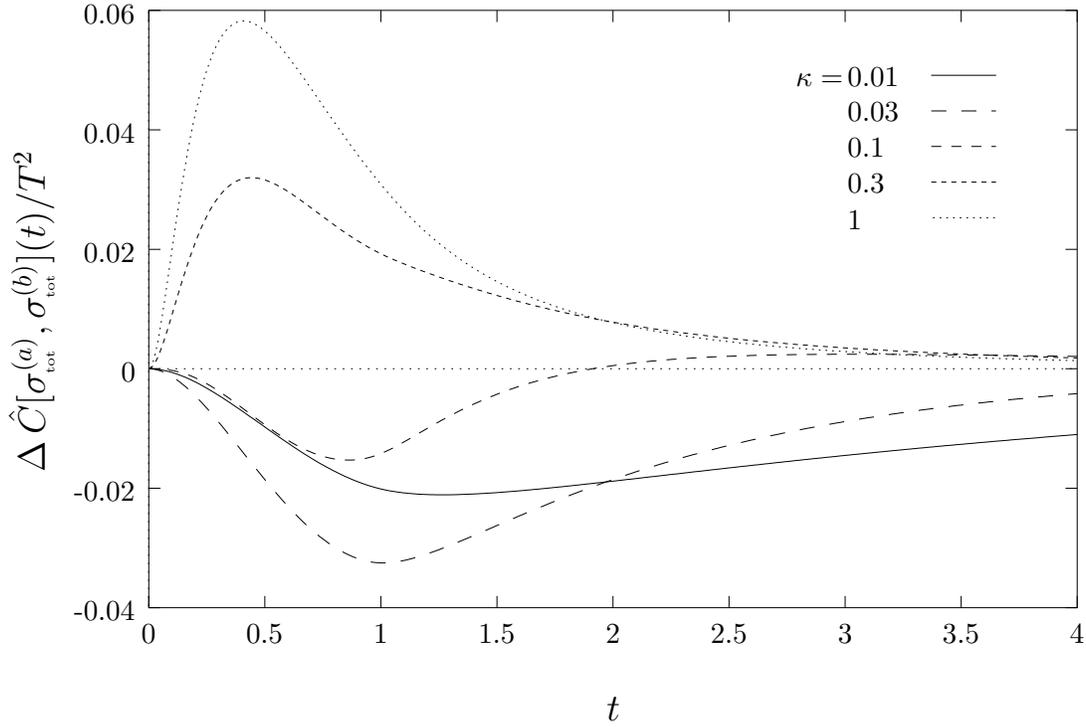}
\caption{As in Fig.~\ref{f_dK2auto}, but for the cross correlation function
$\hat C[\sigma_\tot^a,\sigma_\tot^b](t)$ divided by $T^2$.}
\label{f_dK2cross}\end{figure}

The Figs.~\ref{f_d1-1auto}--\ref{f_dK2cross} show the difference 
between the rescaled Breit-Wigner approximation (\ref{C_FCS-2}) and the 
exact result (\ref{V_FVWZ}) for C-functions of the total cross section 
in the GOE case. This is done for different numbers of channels 
(Fig.~\ref{f_d1-1auto} and Fig.~\ref{f_d1-1cross}), and for different 
coupling strengths (Fig.~\ref{f_dK2auto} and Fig.~\ref{f_dK2cross}). As the 
error of the rescaled Breit-Wigner approximation is zero for $t=0$ (see 
Sec.~\ref{VST}), all curves for the auto and cross C-functions begin at 
zero. In the case of the auto C-function the rescaled Breit-Wigner 
approximation mostly underestimates the exact result. In particular at small
times, we get maximal deviations, which decrease again as $t\to\infty$. In the
case of cross C-functions the error curves depend in a more 
irregular way on the number of channels or on the coupling strength. Though
the error is of comparable size in all cases, the error curves at $t\to 0$
seem to behave linearly in the case of the auto C-functions and quadratically 
in the case of the cross C-functions (in agreement with our theoretical
expectations; see Sec.~\ref{VST}). Yet the cross C-function vanishes for
$\kappa = 1$, while the approximation does not and must, therefore, be considered 
with great reservations. In general, the error is rather an absolute then a 
relative one. We may say that the rescaled Breit-Wigner approximation works 
quite well up to $\kappa =0.1$, and that for larger $\kappa$ the approximation 
still behaves qualitatively well except for the cross C-function for
$\kappa =1$.

\begin{figure}
\input{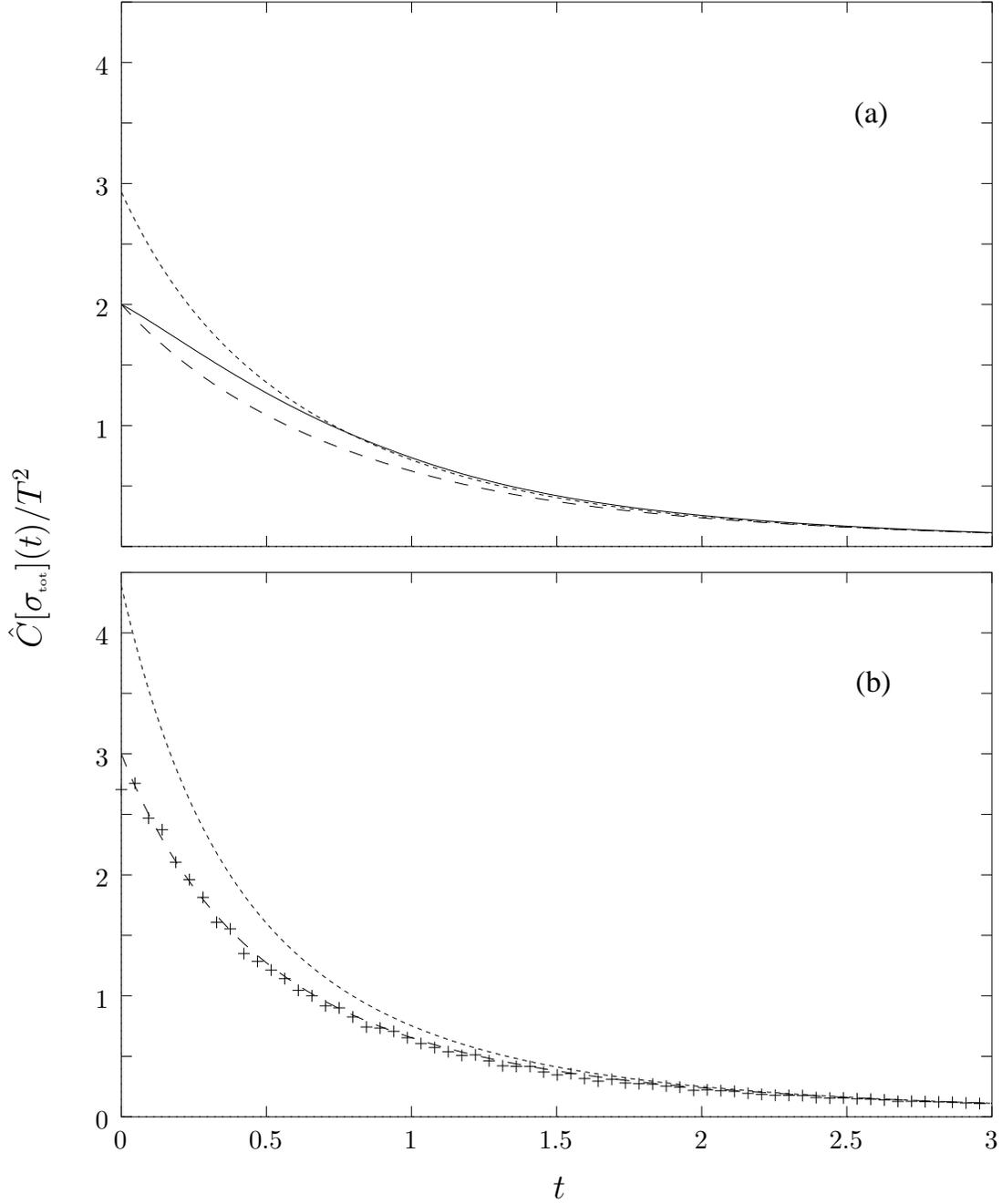}
\caption{The autocorrelation function $\hat C[\sigma_\tot](t)$ divided by $T^2$
for $M = 2$ and $\kappa = 0.1$. The long dashed line is the rescaled
Breit-Wigner approximation (\ref{C_FCS-2}), and the short dashed line is the 
ordinary Breit-Wigner approximation (details see text). In (a) we show the GOE
and in (b) the POE case. The solid line in (a) shows the exact result 
(\ref{V_FVWZ}), whereas the crosses in (b) show numerical Monte Carlo 
calculations.}
\label{f_xxx}\end{figure}

In Fig.~\ref{f_xxx}, we compare the rescaled and
the ordinary Breit-Wigner approximation. The latter is obtained by undoing the
rescaling and replacing $T$ by $4\kappa$ in Eq.~(\ref{C_FCS-2}). As we 
have no exact theory in the POE case [Fig.\ \ref{f_xxx}(b)], we use numerical
Monte Carlo data instead ({\it cf.} Sec.~\ref{N}). The data may be considered as 
exact up to the small fluctuations, seen in the data points. Though the 
resonances are well separated for $\kappa=0.1$ which is used here, the pure 
Breit-Wigner approximation is already far off the exact result, in particular 
at small times. This is true for the GOE case shown in Fig.~\ref{f_xxx}(a), 
as well as for the POE case in Fig.~\ref{f_xxx}(b). In fact, the 
rescaled Breit-Wigner prediction in the GOE case is much closer to the POE
case, which makes it practically impossible to distinguish between correlated
and uncorrelated spectra. Note that the auto C-function shown here, is 
monotonously decreasing without showing any qualitative sign of the correlation 
hole.

The rescaled Breit-Wigner approximation by contrast, reproduces the exact 
values of the C-function at $t=0$ in the GOE and apparently also in the POE 
case. This allows to detect the correlation hole if present, at least in 
principle (absolute cross sections must be available with sufficient accuracy).
Though the rescaled Breit-Wigner approximation underestimates the auto 
C-function at intermediate times in the GOE case [Fig.\ \ref{f_xxx}(a)],
it agrees very nicely with the numerical data in the POE case
[Fig.\ \ref{f_xxx}(b)]. \\

It seems that the rescaled Breit-Wigner accounts very well for 
uncorrelated resonances that diffuse independently into the complex plane 
(similar to what one would expect from an effective mean field theory). For 
resonances which are strongly correlated from the very beginning, the dynamics 
of the resonances is more complicated, and the rescaling procedure cannot 
fully account for that. However, in cases where deviations of a few percents 
(of the maximal C-function value) are still acceptable, the rescaled 
Breit-Wigner approximation can be applied up to quite large coupling strengths,
which reach well into the strong absorption regime. While it is convenient to 
consider the absolute error at small times (in the correlation hole region),
the relative error is more appropriate at large times, {\it i.e.} in the
tails of the C-function. The latter apparently scales with the total 
transmission $\sum_{a=1}^M T_a$ (we have checked this separately), in line with 
earlier theoretical work \cite{Ver86}.

\section{\label{N}Numerical analysis: GOE versus POE}

Here we perform Monte Carlo simulations for correlation functions of partial 
and total cross sections. The aim is to distinguish the situations, where the 
internal Hamiltonian has a purely random spectrum (POE), from those where it
has a spectrum with correlations (GOE). We restrict the analysis to coupling 
strengths $0 < \kappa \le 1$ that covers the whole range for the transmission 
coefficient. As discussed amongst others in
Refs.~\cite{Dit91b,Gor97b,AlbSeb97} the 
scattering ensemble may show quite a different behavior when $\kappa\gg 1$.
The results are presented for two open channels. Calculations for larger 
channel numbers have been performed, but they show no significantly different 
behavior. In the case of total cross sections and GOE, this can be seen from
the results for the VWZ integral obtained in Sec.~{V} (see for example 
Fig.\ \ref{f_C1-1auto}). In the other cases, {\it i.e.} POE or partial cross 
sections, the rescaled Breit-Wigner approximation gives the same answer [see 
Eq.~(\ref{R_FCsigab})]. In all cases the number of 
channels merely enters as a scale factor.

The numerical data are obtained using Eq.~(\ref{C_cthm}), {\it i.e.} we
calculate the Fourier transforms of the two cross sections in question and then 
take the ensemble average ($400$ runs) over their (complex conjugated) product. 
Finally the product of the average cross sections is subtracted to eliminate 
the peak at $t=0$. The dimension of the effective Hamiltonian $H$ [see 
Eq.~(\ref{M_Smat})] is $N=300$. Where the resulting data are smoothed, 
this is done using a $\chi^2$-fit to a natural cubic spline. The length of its 
curve segments was 28 points that corresponds to $\Delta t \approx 0.26$ 
\cite{gnuplot99}. From the spline, $150$ equidistant points in $t$ are finally 
plotted. Note, that there is no free fit parameter, and the curves are not 
normalized. The factor $T^2$ by which we divide, is directly determined from 
the input parameter $\kappa$, using Eq.~(\ref{M_Tau}).

\subsection{Total cross sections}

We start the numerical analysis with auto and cross correlation 
functions of total cross sections. While this case is relatively easy to handle
in theory, it is usually very difficult to obtain total cross sections from an
experiment, in particular if two independent total cross sections are needed.

\subsubsection*{Autocorrelation functions}

As a starting point consider the correlation hole \cite{Lev86} in the Fourier 
transform of a discrete energy spectrum $\hat C(t)\sim 1-b_2(t)$, that 
distinguishes the GOE from a purely random spectrum (POE). In the limit of 
vanishing coupling $\kappa\to 0$, the auto C-function $\hat C[\sigma_\tot](t)$
divided by $T^2$ converges not to $\hat C(t)$, but to $\hat C(t)+2$. The reason 
is that the resonances are weighted by Porter-Thomas distributed intensities 
(or partial widths) \cite{Lom93,AlGuHa97}, which leads to a reduced correlation 
hole of $1/3$. 

\begin{figure}
\input{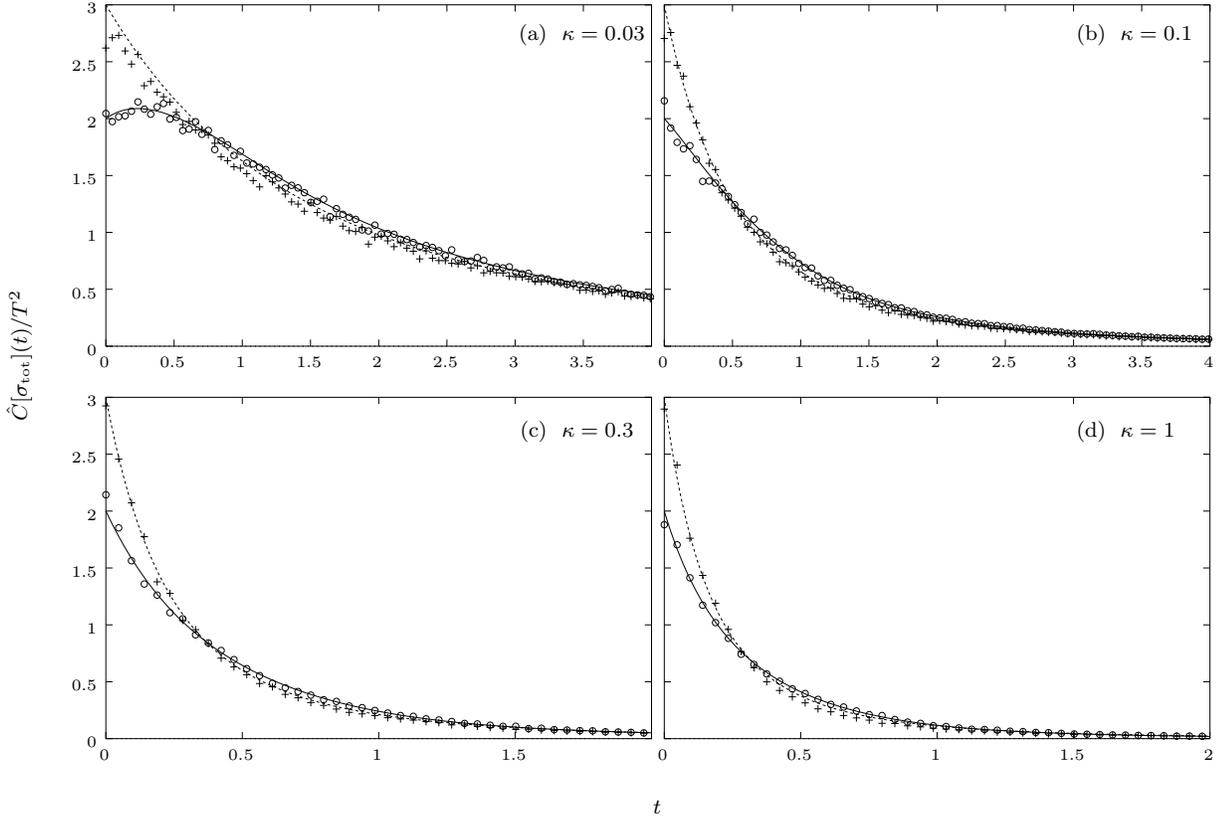}
\caption{The autocorrelation function of the total cross section divided
by $T^2$. The GOE versus the POE case, for coupling strengths: $\kappa = 0.03$
(a), $\kappa = 0.1$ (b), $\kappa = 0.3$ (c), and $\kappa = 1.$ (d). The circles 
show the smoothed numerical GOE data; the crosses the equally treated POE data.
In all cases (a),\ldots,(d) the exact VWZ integral is used for the GOE theory 
(solid line) and the rescaled Breit-Wigner approximation for the POE (dashed 
line).}
\label{f_GP1_22}\end{figure}

Figure~\ref{f_GP1_22} shows $\hat C[\sigma_\tot](t)/T^2$, for the 
GOE and the POE case. We find the behavior described above, but with the
difference, that due to the finite coupling $\hat C(t)$ must be multiplied 
with roughly an exponential decaying function. In the four panels~(a)--(d),
the coupling parameter is increased from $\kappa= 0.03$ to $\kappa= 1$. In the 
GOE case the numerical results are compared with the exact theory, 
Eq.~(\ref{V_FVWZ}), but in the POE case we are dependent on the rescaled 
Breit-Wigner approximation, Eq.~(\ref{C_FCS-2}). We find perfect agreement 
for the GOE which shows that the numerical Monte-Carlo calculation is reliable. 
Interestingly the rescaled Breit-Wigner approximation describes the POE data 
almost as well. Only for $\kappa = 1$ we find minor deviations at intermediate 
times. This reaffirms, that the rescaled Breit-Wigner approximation is best 
suited for cases where the correlations are relatively weak.

\subsubsection*{Cross correlation functions}

Let us again think of a discrete stick spectrum where the levels are weighted
by intensities (or partial widths). In order to calculate the C-function 
between two spectra related to different channels, it is reasonable to assume 
that the positions of the resonances are the same, but the corresponding
intensities are uncorrelated. Then this cross C-function becomes: 
$\hat C(t) \sim 1-b_2(t)$, {\it i.e.} we obtain the full correlation hole
\cite{privLom} (see also the Eqs.~(\ref{C_FCS-1})--(\ref{C_FCS-2}) and the 
discussion following them). For the GOE case, this has been shown numerically in 
Sec.~\ref{VSC}.

\begin{figure}
\input{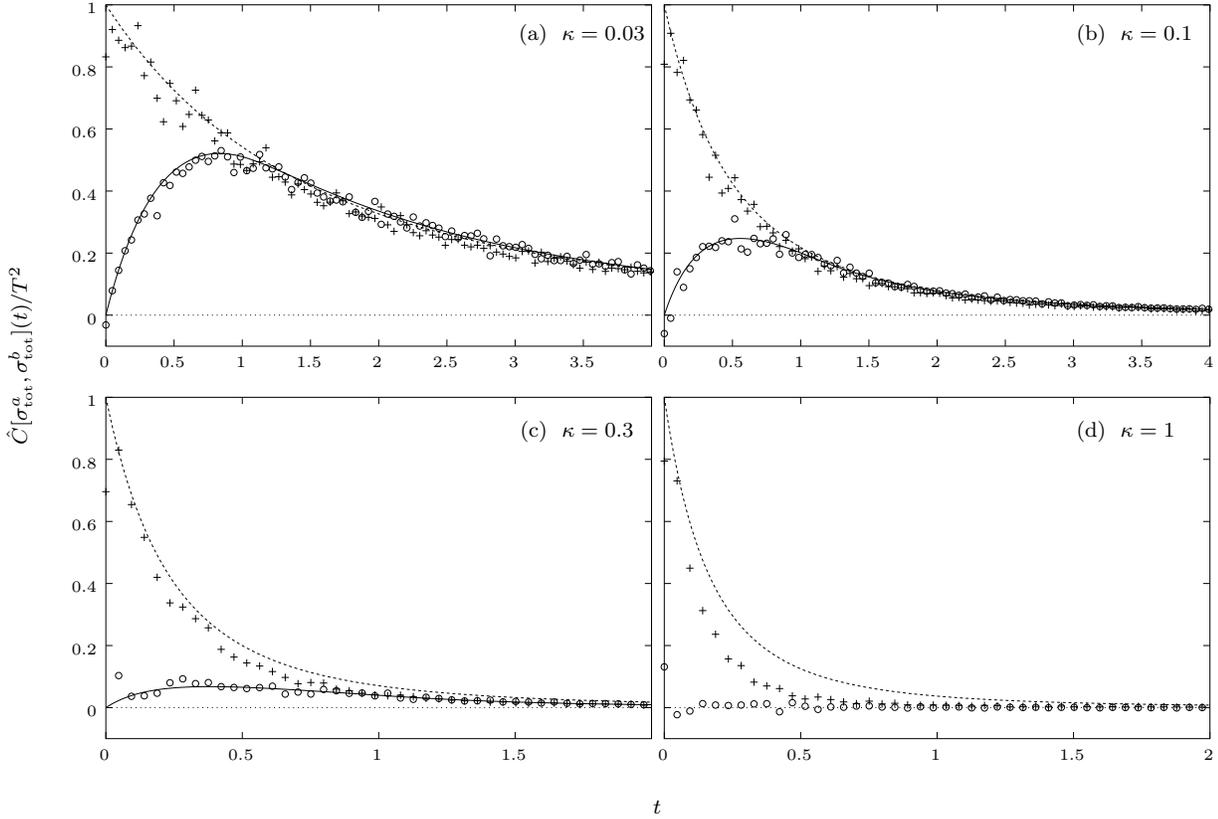}
\caption{As in Fig.~\ref{f_GP1_22}, but for the cross correlation function
divided by $T^2$. Note, however, that in part (d), the C-function in the GOE 
case is exactly zero, and not plotted.}
\label{f_GP1_23}\end{figure}

In Fig.~\ref{f_GP1_23} we show numerical and theoretical results for the 
cross C-function. In the GOE case we plot the numerical results together 
with the exact theory, Eq.~(\ref{V_FVWZ}), and we find again perfect 
agreement well within the statistical error (note that for $\kappa = 1$, the 
cross correlation function vanishes identically). In the POE case we compare 
our numerical data with the rescaled Breit-Wigner approximation, 
Eq.~(\ref{C_FCS-2}). In the panels (a) and (b), which correspond to the 
weak coupling case, the theory agrees very well with the numerical data. For 
$\kappa = 0.3$ (c), however, we find first systematic deviations. The rescaled 
Breit-Wigner approximation begins to overestimate the true C-function, which 
becomes even worse for $\kappa =1$ (d).

Possibly the deviations for $\kappa \gtrsim 0.3$ may be due to the appearance 
of correlations when the resonances begin to overlap. The fact that the 
deviations are more pronounced here, than in the case of auto C-functions can 
be explained in a natural way by the higher sensitivity of the cross C-function 
to correlations. Moreover, it has been proven in Ref~\cite{Gor99} (though only 
for the one channel case) that the coupling to decay channels may indeed induce 
correlations even in the regime $\kappa < 1$. \\

In the GOE case, we know from Sec.~\ref{VST} that the rescaled 
Breit-Wigner approximation reproduces the exact value of auto and cross 
C-functions at $t=0$ for arbitrary coupling strengths. This can also be proved
for the POE in the one channel case \cite{privGor}. Here, Fig.\ \ref{f_GP1_22} 
and Fig.\ \ref{f_GP1_23} give numerical evidence, that the
same is true in the many channel case as well. Thus the relative size of the 
correlation hole seems to be independent of the coupling strengths, which turns 
the correlation hole into a persisting signature of chaos even in the limit of 
strong transmission $\forall a : T_a\to 1$. This may be contrasted to the
behavior in the energy domain, where the correlation hole for 
$C[\sigma_\tot^{(a)},\sigma_\tot^{(b)}](\omega)$ gradually disappears with 
increasing number of channels or total transmission. Note, however, that in the
time domain, the correlation hole becomes compressed to short times. One would 
then need very long spectra with many resonances to resolve the correlation 
hole.  The generality of this invariance property and its implications will be 
the subject of future studies.

\subsection{Partial cross sections}

In this section  we show results of Monte Carlo calculations for C-functions
of partial cross sections. Though we obtained a closed formula even here (using 
the diagonal approximation for the partial cross sections and the rescaled 
Breit-Wigner approximation for the correlation function), we must be prepared 
for the approximation to deviate relatively early from the true result. The 
main advantage of partial cross sections, is their easy experimental 
accessibility. According to our approximation, the relative size of the 
correlation hole is given by:
\begin{equation}
Q= \frac{B}{A}\;\frac{(1+6/M)(1+4/M)}{(1+2/M)^2} 
\end{equation}
(see Eq.~(\ref{R_FCsigab}) and the list below). Note that for particular 
choices of the partial cross sections, this quotient may become even larger 
than one, which implies that the corresponding C-function for the GOE becomes 
negative at small times.

\subsubsection*{Autocorrelation functions}

According to Eq.~(\ref{R_FCsigab}) we have to distinguish between the auto 
C-function of an elastic or inelastic cross section. In the elastic case, 
$B/A= 3/35$ so that the relative size $Q$ of the correlation hole varies from 
$Q= 3/35 \approx 0.0857$ for $M\to\infty$, and $9/35 \approx 0.257$ for $M=2$. 
In the inelastic case, $B/A= 1/9$ so that $Q$ varies from $Q=1/9 \approx 0.111$ 
for $M\to\infty$, and $Q= 1/3 \approx 0.333$ for $M=2$.

\begin{figure}
\input{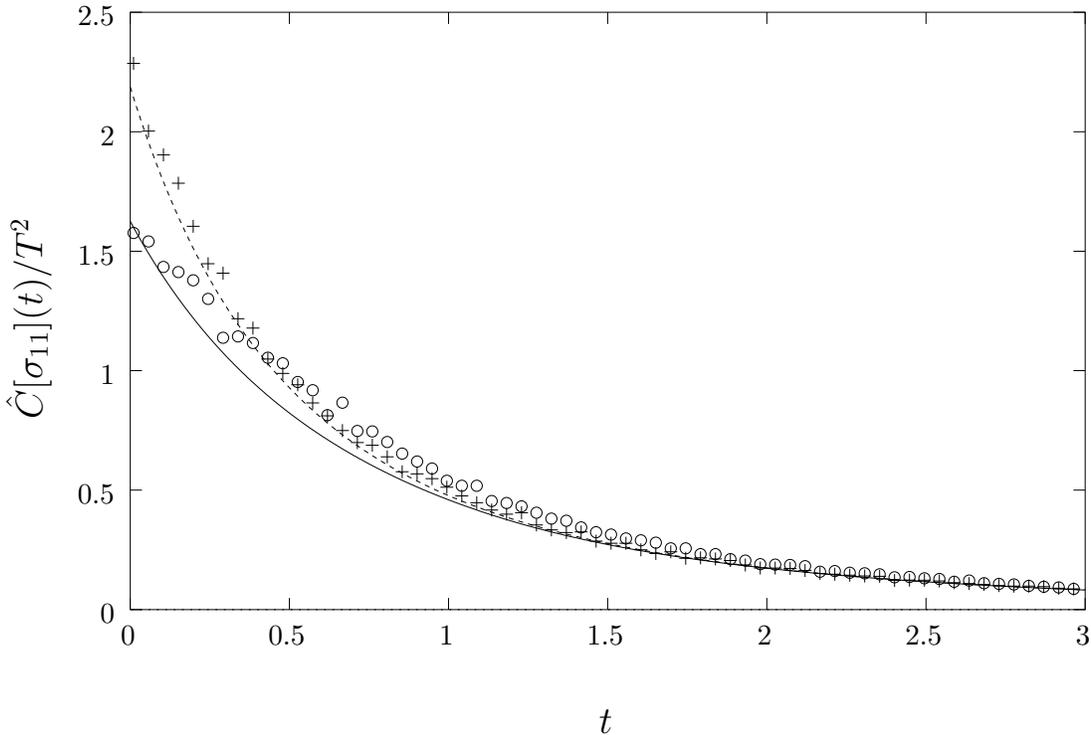}
\caption{The autocorrelation function for the elastic cross section divided
by $T^2$ with the coupling strength $\kappa = 0.1$. The numerical data are 
shown by circles (GOE) and crosses (POE). The theoretical curves are the
rescaled Breit-Wigner approximation, Eq.~(\ref{R_FCsigab}),
for the GOE (solid line) and the POE (dashed line).}
\label{f_GPael24}\end{figure}

\begin{figure}
\input{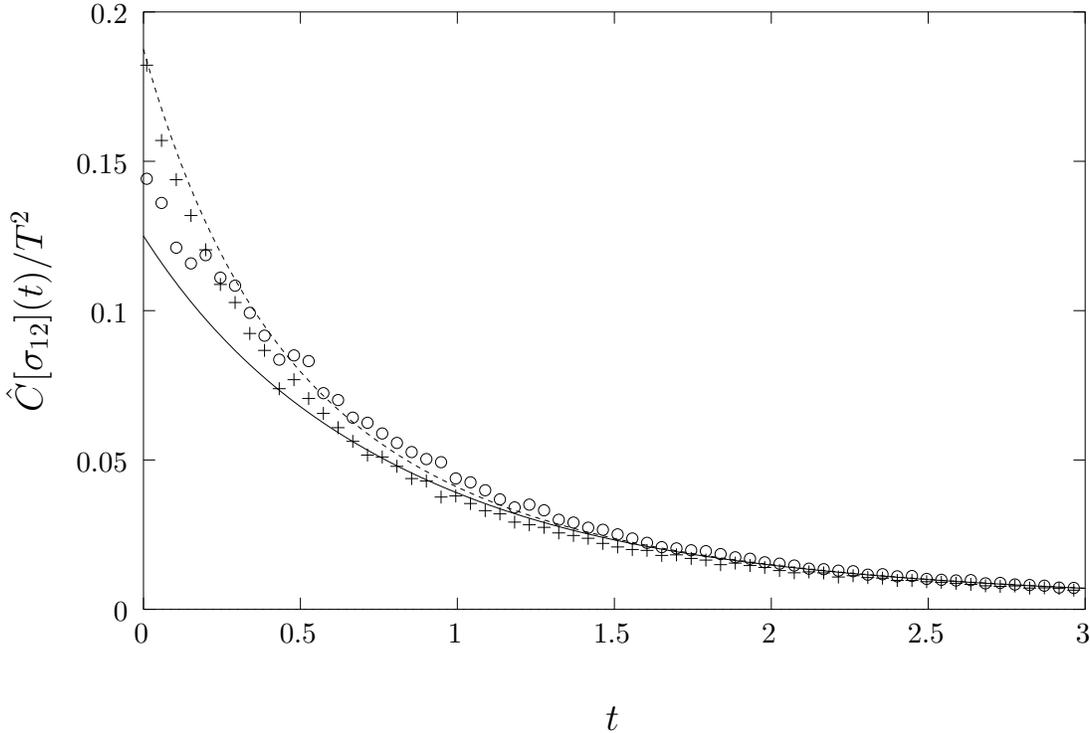}
\caption{As Fig.~\ref{f_GPael24}, but for inelastic cross sections.}
\label{f_GPain26}\end{figure}

In Fig.~\ref{f_GPael24} (elastic case) and Fig.~\ref{f_GPain26} (inelastic 
case) we set the coupling parameter to $\kappa = 0.1$. The numerical data is
shown together with our approximate result, Eq.~(\ref{R_FCsigab}). In both
figures, the approximation clearly deviates from the numerical result 
(which we may consider to be exact). The deviations are more pronounced in the 
GOE case, which is in line with earlier observations in the context of total 
cross sections. The deviations occur just in the region of the correlation 
hole, and thus diminish the difference between GOE and POE considerably. In 
Fig.\ \ref{f_GPael24} it seems, that the theoretical values at $t=0$ are still 
quite accurate, so that the correlation hole could be detected there. However, 
the curves for the GOE and POE merge very rapidly (at $t\approx 0.5$). In 
Fig.\ \ref{f_GPain26} we see that our approximation fails even at $t=0$. The 
numerical curves for POE and GOE lie so close together, that it is practically 
impossible to identify the correlation hole. Similar investigations
\cite{AlGuHa97} gave essentially the same results.

\subsubsection*{Cross correlation functions}

As the number of channels is $M=2$, we need to consider only two different
types of cross C-functions: the C-function of two elastic cross sections, and 
of an elastic and an inelastic cross section (sharing the same entrance 
channel). In the list below Eq.~(\ref{R_FCsigab}) we find that $B/A= 1$ in 
the first case, and $B/A= 1/5$ in the latter. Hence we may expect to obtain 
considerably larger correlation holes. For the case of two elastic cross 
sections the relative size $Q$ of the correlation hole is $Q=1$ for 
$M\to\infty$ and $Q=3$ for $M=2$. This means, that the C-function in the GOE 
case becomes negative at small times. For the case of one elastic and one 
inelastic cross section, $Q=1/5$ for $M\to\infty$ and $Q=3/5$ for $M=2$. This 
is the most natural constellation in experiments. Unfortunately the 
corresponding correlation hole is not very large.

\begin{figure}
\input{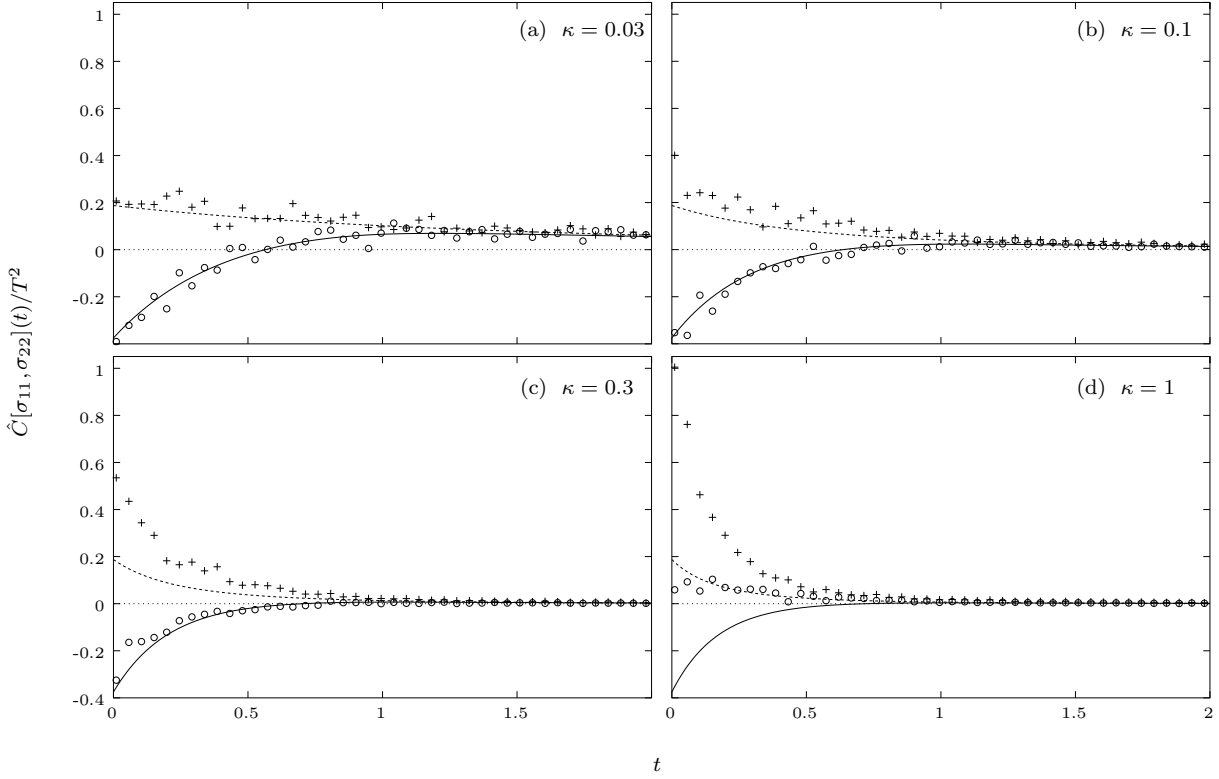}
\caption{The cross correlation function
$\hat C[\sigma_{11},\sigma_{22}](t)$ divided by $T^2$. The coupling strength
is varied: $\kappa = 0.03$ (a), $\kappa = 0.1$ (b), $\kappa = 0.3$ (c), and
$\kappa = 1$ (d). The numerical data is shown with circles (GOE) and crosses
(POE). The rescaled Breit-Wigner approximation (\ref{R_FCsigab}) is
plotted for the GOE (solid line) and for the POE (dashed line).}
\label{f_GP2_28}\end{figure}

Figure~\ref{f_GP2_28} shows the cross C-function of two elastic
cross sections divided by $T^2$. In the four panels (a)--(d) the coupling
strength is increased from $\kappa= 0.03$ up to $\kappa= 1$. The correlation 
hole is so big ($Q=3$), that the GOE curve becomes indeed negative at small 
$t$, with the only exception of panel (d). The rescaled Breit-Wigner 
approximation works quite well in (a) and (b), begins to fail in (c), and fails 
completely in (d). Surprisingly the error becomes large in the POE case first, 
while the agreement in the GOE case is still reasonable. This contradicts to
some extent, what we have found and partially explained in the case of total 
cross sections. Further studies are probably necessary to clarify this point.

\begin{figure}
\input{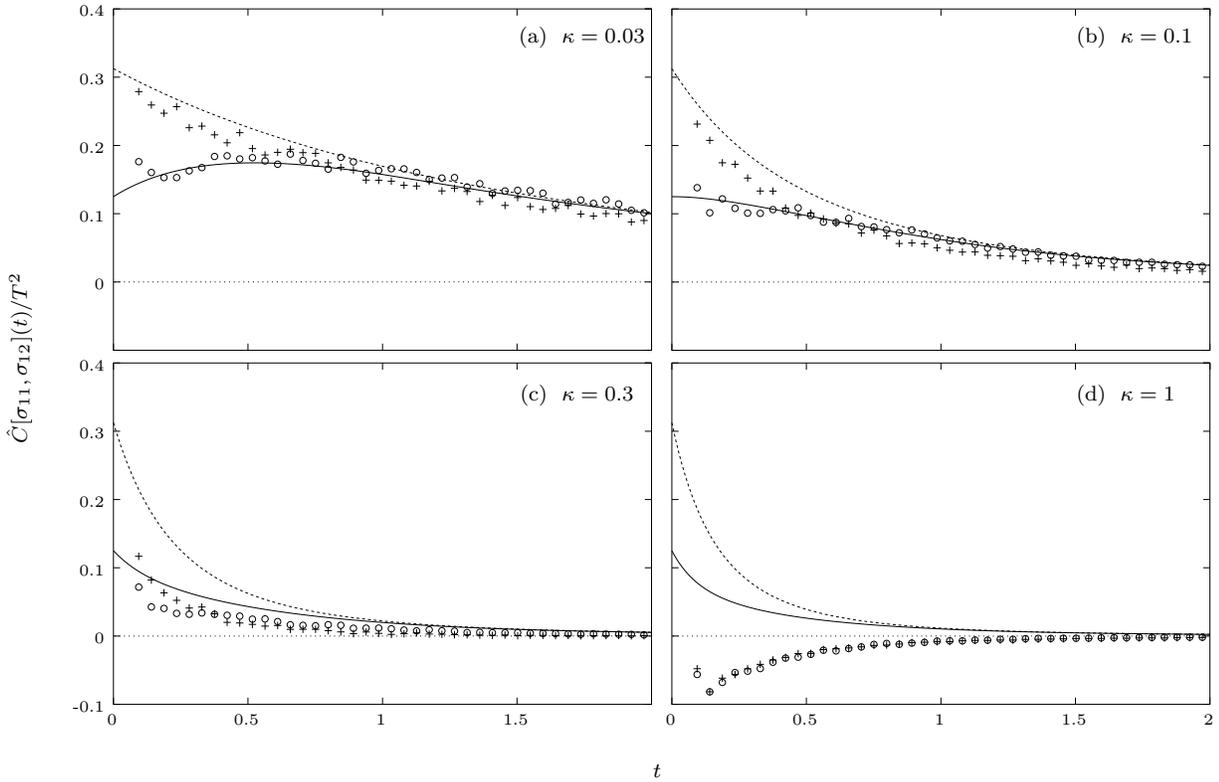}
\caption{As in Fig.~\ref{f_GP2_28}, but for the cross correlation
function between an elastic and an inelastic cross section
$\hat C[\sigma_{11},\sigma_{12}](t)$ divided by $T^2$.}
\label{f_GP2_25}\end{figure}

Figure~\ref{f_GP2_25} shows the cross C-function of an elastic and an inelastic 
cross section. Not only is the correlation hole smaller, the deviations between
our approximation and the numerical data are also larger than in the previous 
case (Fig.~\ref{f_GP2_28}). In panel (b) for $\kappa = 0.1$ we find already 
noticeable deviations in the POE case, and the correlation hole has practically
disappeared for $\kappa =0.3$, in panel (c). In panel (d) finally, the 
numerical data practically coincide for the POE and the GOE case. It seems that
the C-function in this case is negative for any value of $t$. Again this is an
aspect, which deserves further studies. \\

In this subsection, we saw that the validity of our approximation for partial
cross sections is restricted to coupling strengths that are typically much 
smaller than in the case of total cross sections. Moreover, we can no longer 
obtain the exact value of the C-functions at $t=0$. However, we found here the 
largest effects of the correlation hole. In particular, the case of two elastic 
cross sections is very promising in view of a possible realization in an 
experiment.

\section{Summary \label{S}}

By introducing an appropriate rescaling of the coupling strengths in the 
conventional theory of multichannel scattering for weak absorption, we have 
been able to extend its domain of validity far into the region of overlapping 
resonances. We have used this method (which we call the rescaled Breit-Wigner
approximation) to calculate auto and cross correlation functions for total and 
partial cross sections both for the POE and the GOE case, {\it i.e.} in 
situations we consider as ``integrable'' and ``chaotic.''

Starting from the VWZ integral \cite{VWZ85}, which describes the GOE case,
we have performed exact calculations for the correlation functions of total 
cross sections in the time domain. By comparison with these results and 
numerics, we determined the domain of validity of the rescaled Breit-Wigner 
approximation. It turned out, that it works particularly well in the case of
weak correlations.

We showed the differences between POE and GOE all the way up to the domain of 
strong transmission in all channels, and we found that autocorrelation
functions are a poor means of determining the chaoticity or integrability of a 
scattering system in strong absorption, whereas certain cross correlation
functions are quite sensitive. A particularly strong effect is seen in cross
correlation functions of two different total cross sections. Unfortunately 
total cross sections are usually very hard to measure and, therefore, this
observation may be of little practical value. As far as partial cross sections 
are concerned, we find that cross correlations of two elastic cross sections 
are also very sensitive, far more than either the ones between an elastic and 
an inelastic cross section or between two inelastic ones. 

For systems where two asymptotic channels are stable the cross correlation of 
two elastic cross sections is a very attractive means of analyzing the 
reaction. Also at least for intermediate coupling, {\it i.e.} overlapping 
resonances with moderate coupling in each channel we have a theoretical result 
with which to compare. This is particularly important for the integrable case 
where we know that nongeneric behavior might be quite common. First 
experiments on microwave billiards to check our results are on their way.

\section*{Acknowledgments} 

This work was supported by CONACyT grant 25192-E, 
and by DGAPA (UNAM) grant IN-109000. We thank M.~M\"uller for numerical 
calculations, and M.~M\" uller and I.~Rotter for discussions in particular in 
the early stage of this work, as well as Y.~Alhassid for usefull comments.

\begin{appendix}

\section*{\label{aK}Average S-matrix}

The S-matrix given in Eq.~(\ref{M_Smat}) may equivalently be expressed in 
terms of the so called ``K-matrix'' \cite{MahWei69}:
\begin{equation}
S = \frac{1-\rmi K}{1+\rmi K} \; , \quad 
K(E) = \frac{1}{2} V^\dagger \frac{1}{E - H_\intern} V \; .
\label{K_Smat}\end{equation}
As we will see below, the average S-matrix can be obtained easily from the
average K-matrix. This offers a convenient way to define the openness of the 
system, {\it i.e.} the coupling strengths to the decay channels. \\

In the eigenbasis of $H_\intern$ we get the following expression for the 
elements of the K-matrix (\ref{K_Smat}):
\begin{equation}
K_{ab}(E) = \frac{1}{2} \sum_{i=1}^N \frac{V_{ia} V_{ib}}{E-\eps_i} \; ,
\end{equation}
where $\{\eps_i\}$ are the eigenvalues of the closed system $H_\intern$. The 
coupling matrix elements $\{V_{ia}\}$ and the eigenvalues $\{\eps_i\}$ are
statistically independent in the ensembles considered (the GOE
and the POE). Therefore, we get for the average of $K_{ab}(E)$:
\begin{equation}
\la K_{ab}(E^+)\ra = \delta_{ab} \; \frac{N}{2} \; \la V_{ia}\, V_{ib}\ra \;
\lla \frac{1}{E-\eps+\rmi 0}\rra_\rho \; ,
\end{equation}
which holds for any $i$.
Here $\la \ldots\ra_\rho$ stands for the average over the level density $\rho$.
Using that $\la V_{ia} V_{jb}\ra = \delta_{ij}\, \delta_{ab} \la V_{ia}^2\ra$, 
which again holds for the scattering ensembles considered here (see 
Sec.~\ref{M}), we get:
\begin{equation}
\la K_{ab}(E^+)\ra = \delta_{ab}\; \frac{N}{2} \; \la V_{ia}^2\ra \; \left[\;
\pint\rmd\eps \frac{\rho(\eps)}{E-\eps} \; - \; \rmi\pi\, \rho(E) \right] \; ,
\end{equation}
where we split the integral used to average over the level density, into its
principle value part and the residue. Note, that the average K-matrix is 
diagonal. We define the center $E_0 = 0$ of the spectrum as that point where 
the principle value integral vanishes. Then we obtain:
\begin{equation}
\la K_{aa}(0^+)\ra = -\frac{\rmi\pi}{2} \; N\la V_{ia}^2\ra \; \rho(0) \; .
\end{equation}
Finally we define the ``coupling parameters'' $\kappa_a$ as follows:
\begin{equation}
\kappa_a = \rmi \; \la K_{aa}(0^+)\ra = \frac{\pi}{2d}\; \la V_{ia}^2\ra \; , 
\quad d = \frac{1}{N \rho(0)} \; .
\label{K_kap}\end{equation}
It can be shown \cite{Mel85} that due to the analytic properties of the 
S-matrix its average is directly related to the average K-matrix:
\begin{equation}
\la S(E)\ra = \frac{1-\rmi \la K(E^+)\ra}{1+\rmi \la K(E^+)\ra} \; .
\end{equation}
The average S-matrix is also diagonal. In the center of the spectrum the 
average K-matrix is purely imaginary, so that $\la S(E)\ra$ is real. Its 
elements are then given by:
\begin{equation}
\la S_{aa}(0)\ra = \frac{1-\kappa_a}{1+\kappa_a} \; .
\label{SA_sm0}\end{equation}

\end{appendix}

\end{document}